\documentclass[12pt]{iopart}
\usepackage{iopams}  
\usepackage[utf8]{inputenc}
\usepackage[UKenglish]{babel}
\usepackage{graphicx}
\usepackage[squaren]{SIunits}
\usepackage{color}
\usepackage{subfigure}
\usepackage{cite}
\newcommand{\rmr}{\mathrm{r}}
\newcommand{\rml}{\mathrm{l}}

\begin{document}
\title[Dynamics of an active Brownian particle with time-dependent self-propulsion]{Swimming path statistics of an active Brownian particle with time-dependent self-propulsion}

\author{S Babel, B ten Hagen, and H L\"owen}
\address{Institut f\"ur Theoretische Physik II: Weiche Materie, Heinrich-Heine-Universit\"at D\"usseldorf, Universit\"atsstr.~1, D-40225 D\"usseldorf, Germany}
\ead{sbabel@thphy.uni-duesseldorf.de}

\date{\today}

\begin{abstract}
Typically, in the description of active Brownian particles, a constant effective propulsion force is assumed, which is then subjected to fluctuations in orientation and translation, leading to a persistent random walk with an enlarged long-time diffusion coefficient. Here, we generalize previous results for the swimming path statistics to a time-dependent, and thus in many situations more realistic, propulsion which is a prescribed input. We analytically calculate both the noise-free and the noise-averaged trajectories for time-periodic propulsion under the action of an additional torque. In the deterministic case, such an oscillatory microswimmer moves on closed paths that can be highly more complicated than the commonly observed straight lines and circles. When exposed to random fluctuations, the mean trajectories turn out to be self-similar curves which bear the characteristics of their noise-free counterparts. Furthermore, we consider a propulsion force which scales in time $t$ as $\propto \!\! t^\alpha$ (with $\alpha=0,1,2,\ldots$) and analyze the resulting superdiffusive behavior. Our predictions are verifiable for diffusiophoretic artificial microswimmers with prescribed propulsion protocols.
\end{abstract}
\maketitle

\bibliographystyle{unsrt}

\section{Introduction}
The description of self-propelled particles and microswimmers is a rapidly growing domain of statistical physics \cite{catesreview,schimanskyreview,marchettireview}. Even the motion of a single swimmer is non-trivial since this is already a non-equilibrium situation that requires new concepts of statistical mechanics. Due to the micron size of the swimmers, inertial effects are negligible so that the Reynolds number is small, but there are Brownian fluctuations as for passive colloidal particles \cite{ivlev2012complex,Loewen:01,HL_Cyl,Klein}.
In its simplest form, one can generalize the Brownian dynamics of passive colloidal particles to self-propelled particles by including an additional driving term which leads to a constant propagation speed along the particle orientation.
The orientation, however, is subject to thermal fluctuations and therefore there is a non-trivial coupling between particle orientation and translation. For such an active Brownian particle, low-order moments of the time-dependent displacement distribution have been analytically calculated recently \cite{Howse_2007,tenHagen_JPCM}. Moreover, the full displacement probability distribution has been studied in experiment and simulation \cite{Silber-Li}. The established simple picture of a persistent random walk with a persistence generated by the self-propulsion gives a strongly enhanced long-time diffusion constant as compared to a passive particle.
In the noise-free limit, the swimmer moves deterministically with a speed ${\mathbf v}$ along its orientation on a straight line.

These results have been generalized for swimmers that are subjected to an internal additional torque. When fluctuations are neglected, this leads to a motion on circles in two dimensions \cite{vanTeeffelenL2008,Kuemmel:13} and on helical paths in three dimensions \cite{WittkowskiL2012}. The noise-averaged trajectories are a spira mirabilis \cite{vanTeeffelenL2008} in two and a concho-spiral in three dimensions \cite{WittkowskiL2012}. The former was recently confirmed by experiments on asymmetric self-diffusiophoretic swimmers \cite{Kuemmel:13}.

In all of the previous work on active Brownian particles \cite{Wensink_PNAS,WensinkJPCM,KaiserPRE2013,Baskaran_PRL2013,CatesEPL2013,Bialke_PRL2012,Speck_PRL2013},
the effective propulsion was assumed to be independent of time. In this paper, we consider an explicit time dependence of the propagation speed ${\mathbf v}(t)$, which serves as a given input for the Brownian equations of motion. We calculate moments of the displacement probability distribution analytically and thereby generalize results known from previous work \cite{Howse_2007,vanTeeffelenL2008,tenHagen_JPCM}. Our motivation to do so is
threefold: first, real swimmers usually do not move with a constant propagation speed.
In particular, the swimming stroke itself induces variations in time \cite{FriedrichJEB}.
Even in the simple Golestanian three-sphere swimmer \cite{Najafi2004} the net motion is time-dependent, not to speak about swimming strokes in real microorganisms such as Chlamydomonas \cite{GoldsteinScience,FriedrichPRL2012,BennettPRL2013} or larger swimmers as Daphnia \cite{Ordemann2003,Komin2004,Vollmer2006}. A time dependence on the time scale of the individual swimming stroke is typical rather than an exception. In addition, time-dependent propagation can occur on much longer time scales if bacteria are exposed to chemical or light gradients \cite{Hoell2011}. Therefore, most importantly, the model considered in this paper generalizes the previous coarse-grained models with constant self-propulsion towards a more realisitic description of the propulsion mechanism itself. Second, artificial diffusiophoretic microswimmers \cite{Anderson,Ismagilov,2007Kapral,Golestanian_2007,Popescu2010} offer the fascinating possibility of tuning the propagation speed on demand by varying the laser power externally \cite{BechingerJPCM} such that any prescribed form of ${\mathbf v(t)}$ can be programmed and our model is realized.
Third, an analytical solution is interesting in itself as it may serve as a simple test case for experimental and simulation data.

We provide analytical solutions for both the noise-free and the noise-averaged swimming paths for time-periodic propulsion under the action of an additional constant torque. When fluctuations are neglected, such an oscillatory swimmer moves on closed trajectories that can be much more complicated than the commonly observed straight lines and circles. In the presence of translational and rotational Brownian random motion, the mean swimming path turns out to be a \textit{self-similar} curve that still bears the characteristics of the noise-free case under very general periodicity assumptions. Self-similarity is known from many other areas of statistical physics, such as fractals \cite{Peitgen}, growth processes \cite{Meakin_fractals}, networks \cite{Song}, and critical phenomena \cite{Sornette}. Therefore our findings introduce the concept of similarity into the world of mean microswimmer paths. As an example for a non-periodic realization of the self-propulsion, we consider a power-law time dependence and show that a propagation speed which scales in time $t$ as $\propto \! t^\alpha$ (with $\alpha=0,1,2,\ldots$) induces superdiffusive behavior characterized by an exponent $2\alpha +1$ in the time-dependent mean square displacement.

The paper is organized as follows: in section \ref{model} we describe the model equations of active Brownian particles. Results are presented in sections \ref{time-dep} and \ref{time-dep+torque}, where in the latter an additional constant torque is considered on top of the time-dependent self-propulsion. Finally, we conclude in section \ref{summary}.

\section{The model}\label{model}
In our model we consider colloidal particles in a dilute solution where particle--particle interactions can be neglected. The dynamics in the low-Reynolds number regime is governed by the Langevin equations for completely over-damped Brownian motion. We assume that the motion of the particles is constrained to a two-dimensional plane. Such a situation is often realized in experiments with microswimmers  
where gravity keeps them close to the substrate \cite{theurkauff2012dynamic,Silber-Li,Kuemmel:13}.
However, the generalization to three dimensions is straightforward when following the procedure presented in reference \cite{tenHagen_JPCM} for self-propelled particles with constant propagation speed.
 
\begin{figure}[tbh]
\centering
\subfigure[]{\includegraphics[width=0.45\textwidth]{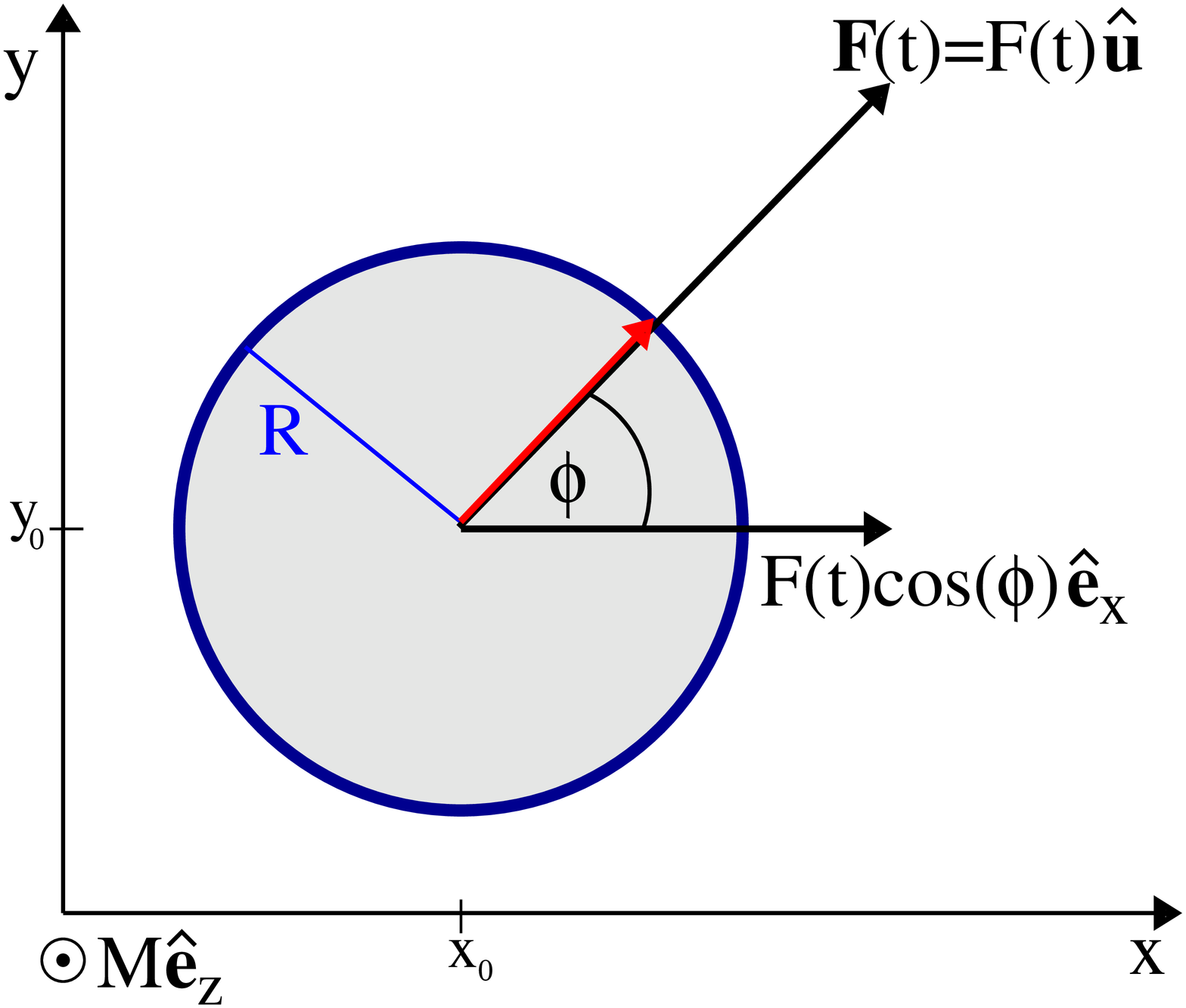}}
\qquad
\subfigure[]{\includegraphics[width=0.45\textwidth]{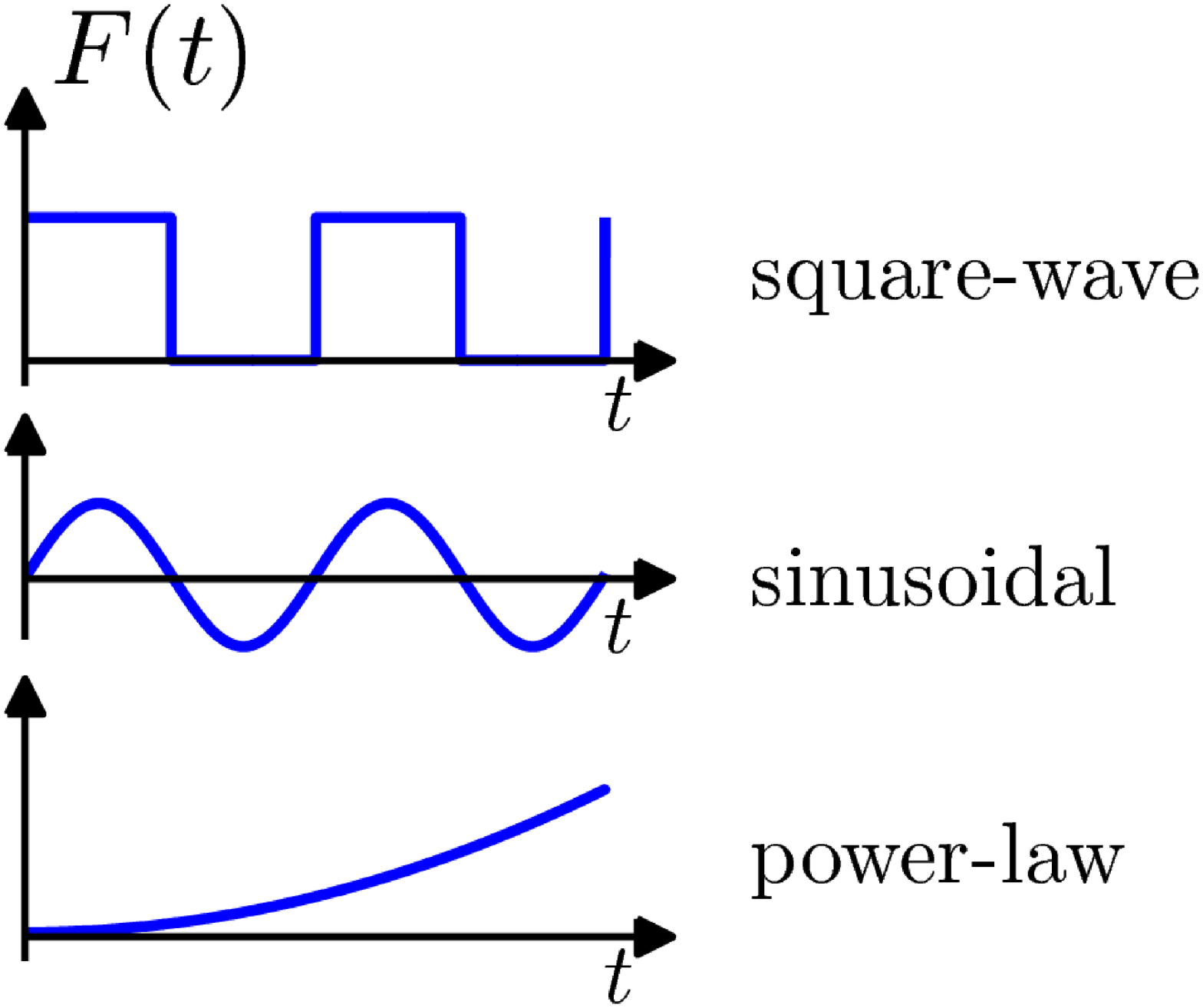}}
\caption{(a) Schematic view of a spherical self-propelled particle with hydrodynamic radius $R$ as considered in our model. The motion is restricted to the two-dimensional $x$-$y$ plane and characterized by the center-of-mass position $\mathbf{r}(t)=[x(t),y(t)]$ and the angle $\phi(t)$ representing the orientation $\hat{\mathbf{u}}=(\cos\phi,\sin\phi)$ of the particle relative to the $x$ direction. The propulsion speed is determined by an effective time-dependent driving force $\mathbf{F}(t)=F(t)\hat{\mathbf{u}}$ and the particle may additionally be exposed to a constant torque $\mathbf{M}=M\hat{\mathbf{e}}_z$ (see section \ref{sec:torque}). (b) Overview of the types of self-propulsion with square-wave, sinusoidal, and power-law time dependence that are explicitly considered.\label{fig:model}}
\end{figure}
 
The colloid itself is regarded as a sphere with a hydrodynamic radius $R$. Its swimming path is determined by the center-of-mass position $\mathbf{r}(t)=[x(t),y(t)]$. To account for the detailed self-propulsion mechanism, we consider an effective time-dependent driving force $\mathbf{F}(t)=F(t)\hat{\mathbf{u}}$, where $\hat{\mathbf{u}}=(\cos\phi,\sin\phi)$ is a particle-fixed orientation vector defined by the angle $\phi$ between the $x$ axis and the direction of propulsion (see figure \ref{fig:model}(a)). Thus, the corresponding translational and orientational Langevin equations are given by  
\begin{eqnarray}
\frac{dx(t)}{dt}&=\beta D \left[F(t)\cos(\phi(t))+f_x(t)\right]\label{1x}\,,\\
\frac{dy(t)}{dt}&=\beta D \left[F(t)\sin(\phi(t))+f_y(t)\right]\label{1y}\,,\\
\frac{d\phi(t)}{dt}&=\beta D_\rmr \, g(t)\label{2} 
\end{eqnarray}
with the inverse effective thermal energy $\beta=1/(k_\mathrm{B} T)$. Brownian random fluctuations are implemented in equations (\ref{1x})-(\ref{2}) by means of zero-mean Gaussian noise terms $f_x(t)$, $f_y(t)$, and $g(t)$. The respective variances are given by 
$\langle f_x(t)f_x(t') \rangle=\langle f_y(t)f_y(t') \rangle=2\delta(t-t')/(\beta^2 D)$ and $\langle g(t)g(t')\rangle =2\delta(t-t')/(\beta^2 D_\rmr)$, where angular brackets denote a noise average. The translational and rotational Brownian motion is characterized by the respective short-time diffusion constants $D$ and $D_\rmr$ fulfilling $D/D_\rmr=4R^2/3$ for a spherical particle. 
As equations (\ref{1x}) and (\ref{1y}) for the motion in $x$ and $y$ direction are formally identical for changed initial conditions, we will only present the results for the $x$ component, but discuss trajectories in the full $x$-$y$ plane. 
To solve the system of Langevin equations (\ref{1x})-(\ref{2}), first the angular equation (\ref{2}) is considered. As the noise term $g(t)$ is Gaussian, following Wick's theorem the full angular probability distribution has to be Gaussian as well and can be obtained by calculating the first two moments of $\phi(t)$ (for more details see references \cite{Doi_Edwards_book,Cond_Matt}). Using the orientational probability distribution as an input for the translational Langevin equations, analytical results for the mean position and the mean square displacement can be derived.

To account for the variable propagation speed which is often observed in the motion of real microswimmers, we study the influence of different types of time-dependent driving forces $F(t)$. Explicitly, we consider piecewise constant, sinusoidal, and power-law realizations of the self-propulsion (see figure \ref{fig:model}(b)). A piecewise constant or ``square-wave'' self-propulsion force (see section \ref{swave}) can mimic biological microorganisms which undergo a run-and-tumble motion \cite{Berg:72,CatesPRL2008,GoldsteinScience,NashPRL}, for example. 
When the swimming stroke itself leads to periodic variations in the propagation speed, a continuous description such as the sinusoidal driving force (see section \ref{sinusoidal}) is the most appropriate one. 
Finally, a power-law type of self-propulsion (see section \ref{power}) may be relevant for organisms that enhance their swimming velocity by consuming food \cite{Strefler:09} or in situations where the velocity of a predator is determined by the prey gradient \cite{Arditi2001}. 
Furthermore, growing clusters of active particles \cite{Cremer2013} may require a power-law time dependence for the description of the propulsion.

Whereas some realizations of the different self-propulsion types can be directly studied experimentally with active particles in nature, such as the run-and-tumble motion of biological microorganisms \cite{Berg:72,Fletcher2014}, recent progress in the field of artificial colloidal microswimmers makes it possible to tune man-made self-propelled objects in a way such that all considered kinds of swimming behavior are realized. This can be accomplished either by an external magnetic field \cite{Baraban_ACSnano} or in systems where the self-propulsion mechanism is triggered by a light source which can be switched on and off \cite{PalacciScience} or be regulated in a more sophisticated way \cite{VolpeSM1,BechingerJPCM}. Thus, any propulsion protocol can be achieved for diffusiophoretic artificial microswimmers.

\section{Time-dependent self-propulsion}\label{time-dep}
\subsection{Square-wave self-propulsion force}
\label{swave}
First, we discuss the mean position and the mean square displacement of a particle propelling through a liquid as governed by the square-wave self-propulsion force  
\begin{equation}
F(t) = \cases{
F_0 &for $nT< t \le (n+\frac{1}{2})T$ \\
0 &for $(n+\frac{1}{2})T< t \le (n+1)T$\\}\\
\textnormal{with } n=0,1,2,...\,, 
\label{squarewave}
\end{equation}
where $T$ ist the cycle duration (see inset in figure \ref{fig:mpsquarewave}).
Active and passive time intervals of equal length alternate. Here, we consider the case of a particle starting with the active regime (constant self-propulsion force $F_0$).
Some of the statements below have to be modified if the particle starts in the exclusively diffusive regime.  

\subsubsection{Mean position}
As a result, the one-dimensional mean position of a particle with the self-propulsion force as defined in equation (\ref{squarewave}) is given by 
\begin{eqnarray}
\label{eq:mpsquarewave}
\fl \langle x(t)-x_0\rangle = \frac{4}{3}\beta F_0 R^2\cos(\phi_0) \nonumber \\
\fl \qquad  \times
\cases{
\parbox{7cm}{\begin{eqnarray}
\fl\left[\frac{\rme^{D_\rmr T}-\rme^{-D_\rmr (n-1)T}}{\rme^{D_\rmr T }+\rme^{D_\rmr T/2}}
+\rme^{-D_\rmr nT}-\rme^{-D_\rmr t}\right] \nonumber
\end{eqnarray}}
 &  \parbox{5.1cm}{for \mbox{$nT<t\le(n+\frac{1}{2})T$} \\ and $n=0,1,2,\ldots$}\\
 \parbox{7cm}{\begin{eqnarray}
\fl \left[\frac{\rme^{D_\rmr T}-\rme^{-D_\rmr nT}}{\rme^{D_\rmr T }+\rme^{D_\rmr T/2}}\right] \nonumber \end{eqnarray}}
 &\parbox{5.1cm}{for \mbox{$(n+\frac{1}{2})T<t\le(n+1)T$} \\ and $n=0,1,2,\ldots\,\,.$}}
\end{eqnarray}
\begin{figure}[tbh]\centering
\centering
\includegraphics[width=0.75\textwidth]{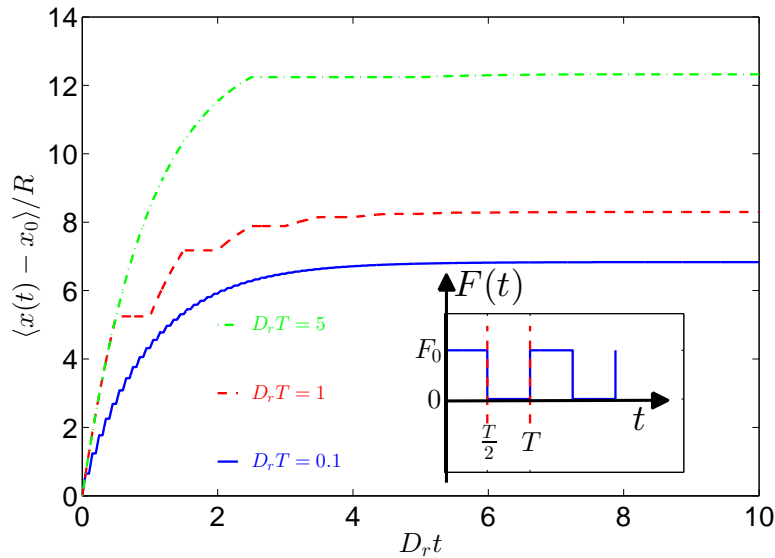}
\caption{Mean position of a self-propelled particle with a square-wave propulsion force based on the analytical result in equation (\ref{eq:mpsquarewave}). Curves are shown for different values of the period $T$ and fixed parameters $\beta R F_0=10$ and  $\phi_0=0$. The stair-like form is due to the lack of an active contribution during every second time interval of length $T/2$ (as visualized in the sketch of the square-wave self-propulsion force in the inset). For long times all curves approach a maximum value which depends on the period length T.\label{fig:mpsquarewave}}
\end{figure}
Obviously, the mean position increases during a time interval of length $T/2$ and stays constant during the following time interval of the same length. This is also visualized in figure \ref{fig:mpsquarewave}, where the dimensionless mean position $\langle x(t)-x_0\rangle/R$ is shown for different values of the scaled period $D_\rmr T$. In all cases the curves exhibit a stair-like form, where the steps are smaller for larger times, and approach a constant value for long times $t$. This final mean position depends on the period $T$ and is obtained as the asymptotic solution from equation (\ref{eq:mpsquarewave}): 
\begin{eqnarray}
\lim\limits_{t,n\rightarrow\infty}\langle x(t)-x_0\rangle=\frac{4}{3}\beta F_0 R^2\cos(\phi_0)\frac{1}{1+{\rme^{-D_\rmr T/2}}}\,.
\label{eq:finalmp}
\end{eqnarray}
Explicitly, the limits $t\rightarrow \infty$ for very short ($T\rightarrow 0$) and very long ($T\rightarrow \infty$) periods are given by 
\begin{equation}
\lim\limits_{T\rightarrow 0}\ \lim\limits_{t,n\rightarrow\infty}\langle x(t)-x_0\rangle=\frac{2}{3}\beta F_0 R^2\cos(\phi_0)
\label{eq:T0}
\end{equation}
and 
\begin{equation}
\lim\limits_{T\rightarrow \infty}\ \lim\limits_{t,n\rightarrow\infty}\langle x(t)-x_0\rangle=\frac{4}{3}\beta F_0 R^2\cos(\phi_0)\,,
\label{eq:Tinfty}
\end{equation}
respectively. Clearly, the result in equation (\ref{eq:T0}) equals the case of a constant self-propulsion force $F=F_0/2=\langle F(t)\rangle=(1/T)\int_{0}^{T}F(t)\mathrm{d}t$ and equation (\ref{eq:Tinfty}) corresponds to the case of a constant self-propulsion force $F_0$.

\subsubsection{Mean square displacement}
While the mean position already elucidates some of the physics of microswimmers with time-dependent self-propulsion, usually the standard quantitiy for characterizing the particle dynamics is the mean square displacement $\langle(x(t)-x_0)^2\rangle$.\\
Our analytical result is as follows: 
\begin{eqnarray}
\fl\langle(x(t)-x_0)^2\rangle=&2Dt +\frac{16}{9}\Big(\beta F_0R^2\Big)^2\Bigg[
D_\rmr \! \left(t-n\frac{T}{2}\right)-n \, \xi\left(\frac{1}{2}\right)-\frac{\cos(2\phi_0)}{12}\ \xi\left(2\right)\rho_{n-1}(4)\nonumber\\
&+\frac{\cos(2\phi_0)}{3}\ \xi\left(\frac{1}{2}\right)\rho_{n-1}(4)+(n-1)\ \xi\left(\frac{1}{2}\right)\rho_{-1/2}(1) \rme^{-D_\rmr T/2}\nonumber\\
&+\xi\left(\frac{1}{2}\right)\rho_{-1/2}(1)\ \rho_{-n}(1)\rme^{-D_\rmr \left(n+1/2\right)T}\nonumber\\
&+\frac{\cos\left(2\phi_0\right)}{3}\ \xi\left(\frac{1}{2}\right)\rho_{-1/2}(3)\Big(\rho_{n-1}(1)-\rho_{n-1}(4)\Big)\nonumber\\
&-\rme^{-D_\rmr T/2}\ \xi\left(\frac{1}{2}\right)\rho_{n-1}(1)\rme^{D_\rmr nT}\ \tilde{\xi}(1)
-\frac{\cos\left(2\phi_0\right)}{3}\ \xi\left(\frac{3}{2}\right)\rho_{n-1}(3)\ \tilde{\xi}(1)\nonumber\\
&+\rme^{D_\rmr nT}\ \tilde{\xi}(1)+\frac{\cos\left(2\phi_0\right)}{12}\ \tilde{\xi}(4)-\frac{\cos\left(2\phi_0\right)}{3}\rme^{-3D_\rmr nt}\ \tilde{\xi}(1)
\Bigg]
\end{eqnarray}
for $nT<t\le(n+\frac{1}{2})T$ with $n=0,1,2,\ldots$ and, correspondingly, 
\begin{eqnarray}
\fl\langle(x(t)-x_0)^2\rangle=&2Dt +\frac{16}{9}\Big(\beta F_0R^2\Big)^2\Bigg[\left(
\frac{D_\rmr T}{2} -\xi\left(\frac{1}{2}\right)\right) \left(n+1\right)-\frac{\cos\left(2\phi_0\right)}{12}\ \xi\left(2\right)\ \rho_n(4) \nonumber\\
&
+\frac{\cos\left(2\phi_0\right)}{3}\ \xi\left(\frac{1}{2}\right)\ \rho_n\left(4\right)+n \, \xi\left(\frac{1}{2}\right)\ \rho_{-1/2}(1) \rme^{-D_\rmr T/2} \nonumber\\
&+\xi\left(\frac{1}{2}\right)\rme^{-D_\rmr \left(n+3/2\right)T}\ \rho_{-1/2}(1)\ \rho_{-\left(n+1\right)}(1)\nonumber\\
&+\frac{\cos\left(2\phi_0\right)}{3}\ \xi\left(\frac{1}{2}\right)\ \rho_{-1/2}(3)\Big(\rho_n(1)-\rho_n(4)\Big)\Bigg]
\end{eqnarray}
for $(n+\frac{1}{2})T<t\le(n+1)T$. Here, the notations 
\begin{eqnarray}
\rho_m(a)&:=\frac{\rme^{aD_\rmr T}-\rme^{-aD_\rmr Tm}}{\rme^{aD_\rmr T}-1}\,,\\
\xi(a)&:=1-\rme^{-aD_\rmr T}\,,\\
\tilde{\xi}(a)&:=\rme^{-aD_\rmr t}-\rme^{-aD_\rmr Tn}
\end{eqnarray}
are used. 

\begin{figure}[tbh]
\centering
\subfigure[]{\includegraphics[width=0.45\textwidth]{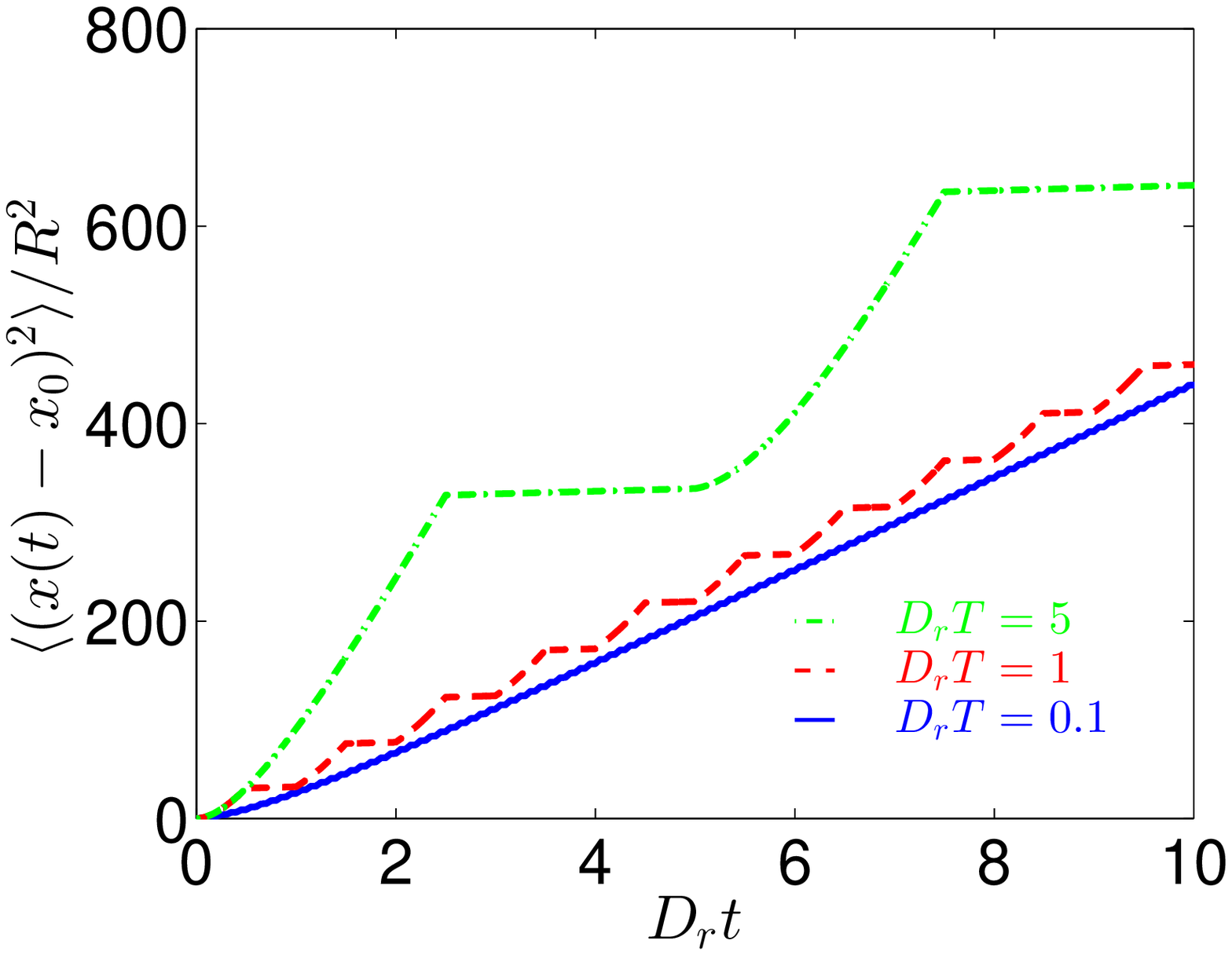}}
\qquad
\subfigure[]{\includegraphics[width=0.45\textwidth]{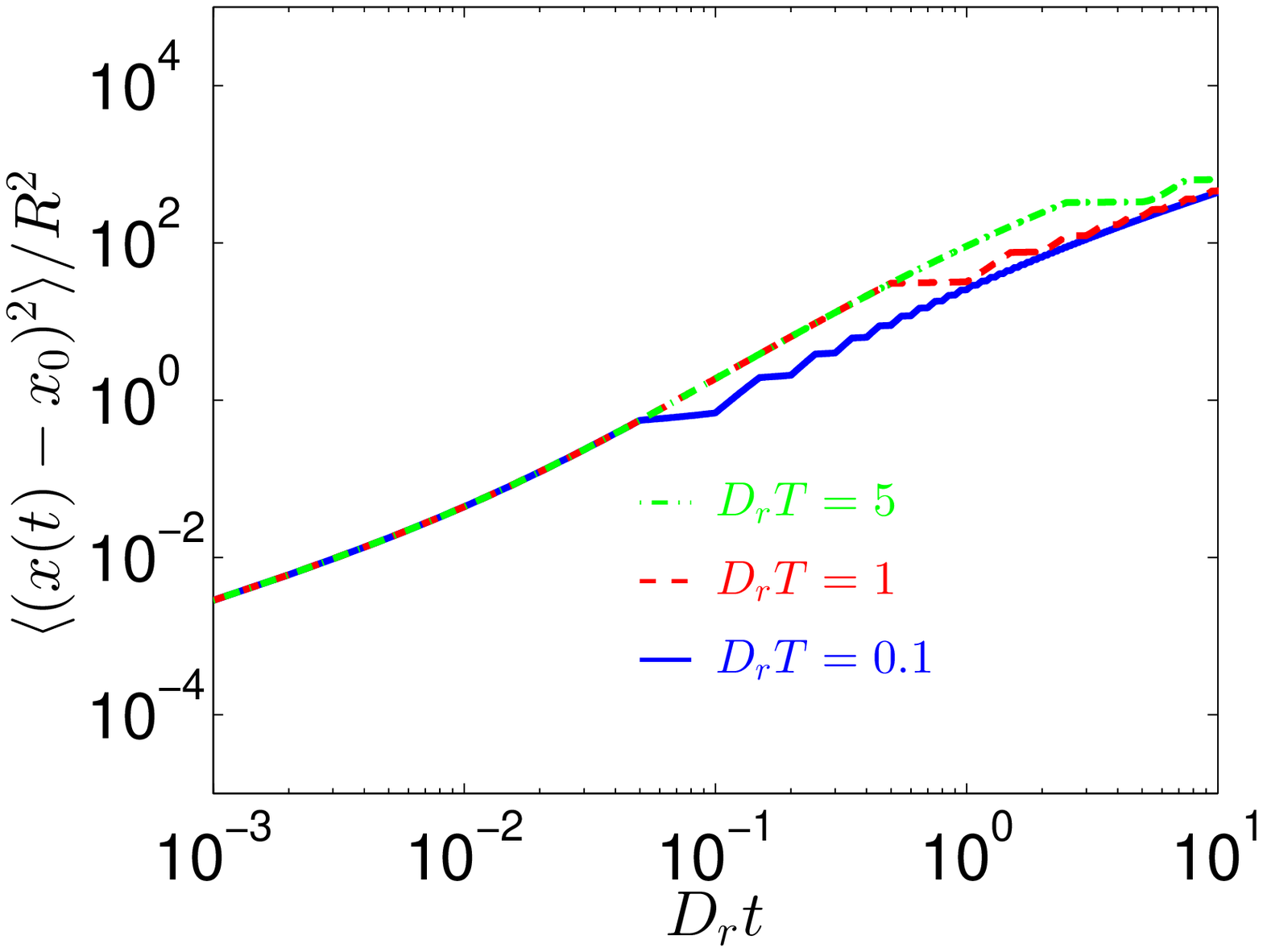}}
\caption{Analytically obtained mean square displacement of a self-propelled particle with a square-wave propulsion force: (a) linear and (b) logarithmic representation for the same situations as in figure \ref{fig:mpsquarewave}. A longer period $T$ leads to larger values of the mean square displacement.}
\label{fig:msdsquarewave}
\end{figure}

Figure \ref{fig:msdsquarewave} visualizes the mean square displacement for the same parameter com\-bi\-nations as in figure \ref{fig:mpsquarewave} for the mean position. The dominant feature is the stair-like pattern resulting from the square-wave force. As shown by the linear representation in figure \ref{fig:msdsquarewave}(a), the steps are significantly more equally sized than with regard to the mean position, where the steps become rapidly flatter with increasing time. The transition from the first active to the first passive regime is most obvious in the logarithmic representation in figure \ref{fig:msdsquarewave}(b). 
In general, a longer period $T$ leads to larger values of the mean square displacement for intermediate and long times. The long-term diffusion coefficient $D_\rml $ for $t\gg T$ is analytically calculated as 
\begin{eqnarray}
\fl D_\rml  & =\lim\limits_{t\rightarrow\infty}\frac{1}{2t}\langle(x(t)-x_0)^2\rangle\nonumber\\
\fl  & =D+\frac{4}{9}\Big(\beta F_0 R^2\Big)^2D_\rmr 
+\frac{8}{9}\Big(\beta F_0 R^2\Big)^2\ \frac{1}{T}\  \frac{\rme^{-D_\rmr T/2}-1}{1+\rme^{-D_\rmr T/2}}\,.
\label{eq:Dlsw}
\end{eqnarray}
For a very short period $T$, equation (\ref{eq:Dlsw}) reduces to 
\begin{equation}
\lim\limits_{T\rightarrow 0}D_\rml =D+\frac{2}{9}\Big(\beta F_0 R^2\Big)^2D_\rmr =D_\rml \Big\vert_{F=\langle F(t) \rangle=F_0/2=\mathrm{const.}}\,,
\end{equation}
which corresponds to the case of a constant self-propulsion force $F= F_0 / 2$. On the other hand, the limit for a very long period is 
\begin{equation}
\lim\limits_{T\rightarrow \infty}D_\rml =D+\frac{4}{9}\Big(\beta F_0 R^2\Big)^2D_\rmr  \,.
\end{equation}
This result exhibits a factor $1/2$ in the second term as compared to the solution for a constant propulsion force $F_0$, which originates from the linear time dependence of the mean square displacement. During every second time interval of length $T/2$ the particle motion is completely passive so that no contribution resulting from the self-propulsion arises. 

\subsection{Sinusoidal self-propulsion force}
\label{sinusoidal}
To account for the effect of a continuous time-periodic propulsion, as often induced by the detailed swimming mechanism of biological microorganisms, we solve the Langevin equations (\ref{1x})-(\ref{2}) for a sinusoidal self-propulsion force 
\begin{equation}
F(t) = F_0(\sin (\omega t)+c) \,.
\label{eq:Fsin}
\end{equation}
It is characterized by the amplitude $F_0$, the frequency $\omega$, and the offset $cF_0$ (see inset in figure \ref{fig4}(a)).

\subsubsection{Mean position}
We obtain
\begin{eqnarray}
\fl\langle x(t)-x_0\rangle=&\frac{4}{3}\beta F_0 R^2 D_\rmr \cos(\phi_0)\Bigg[\frac{\rme^{-D_\rmr  t}}{D_\rmr ^2+\omega^2}\left(-D_\rmr  \sin(\omega t)-\omega\cos(\omega t) \right)\nonumber\\
&+\frac{\omega}{D_\rmr ^2+\omega^2}+\frac{c}{D_\rmr }\left(1-\rme^{-D_\rmr  t}\right)\Bigg]
\label{eq:mpsin}
\end{eqnarray} 
for the particle's mean position. The periodicity resulting from the driving force is washed out for long times. The last term in equation (\ref{eq:mpsin}) vanishes if no constant contribution is considered in equation (\ref{eq:Fsin}), i.e., if $c=0$. In the limit $\omega\rightarrow 0$, as well as for $\omega\rightarrow \infty$, the mean position equals the solution obtained for a constant self-propulsion force $F=c F_0$.

\begin{figure}[tbh]
\centering
\subfigure[]
{\includegraphics[width=0.45\textwidth]{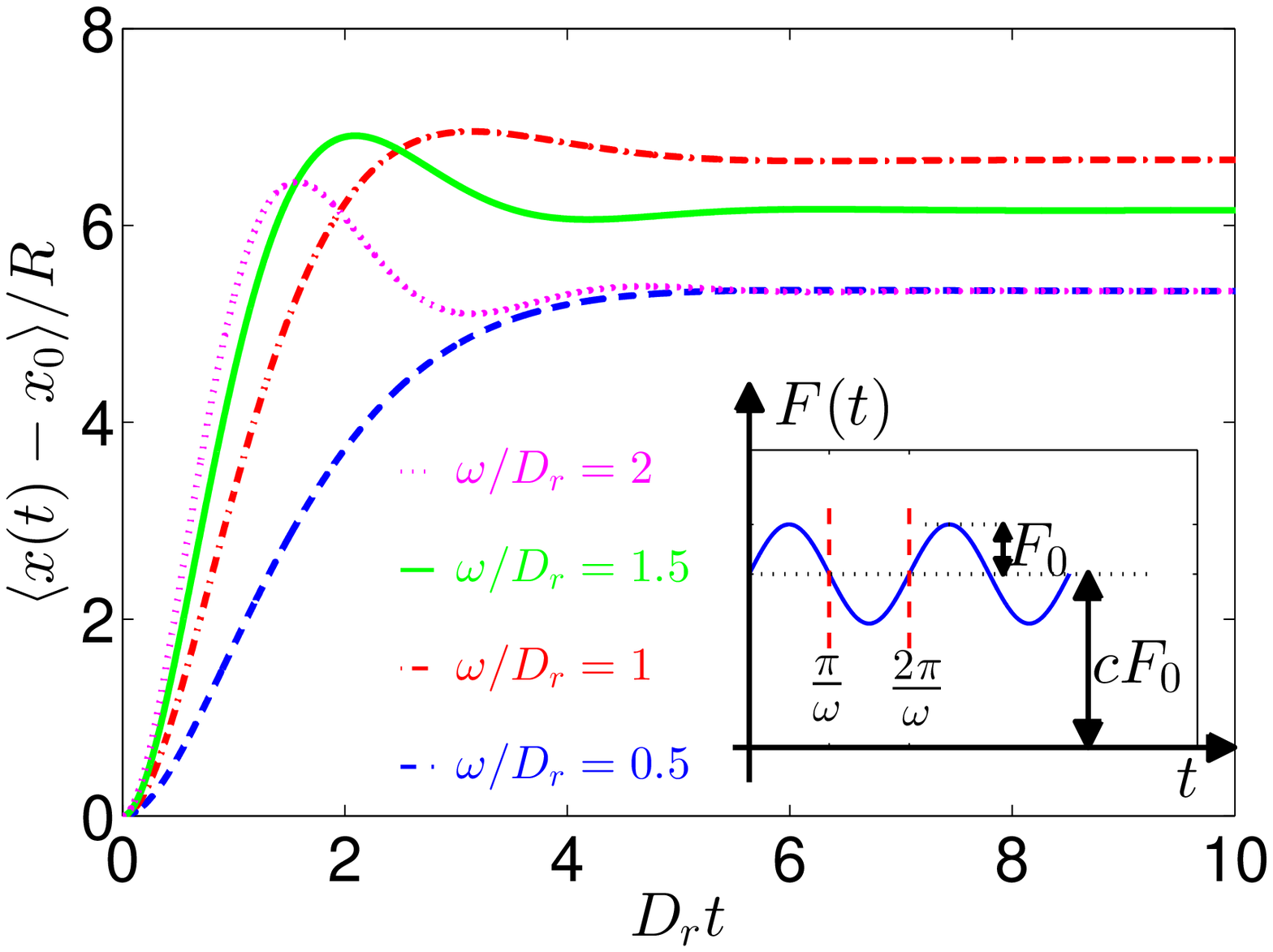}}
\qquad
\subfigure[]{\includegraphics[width=0.45\textwidth]{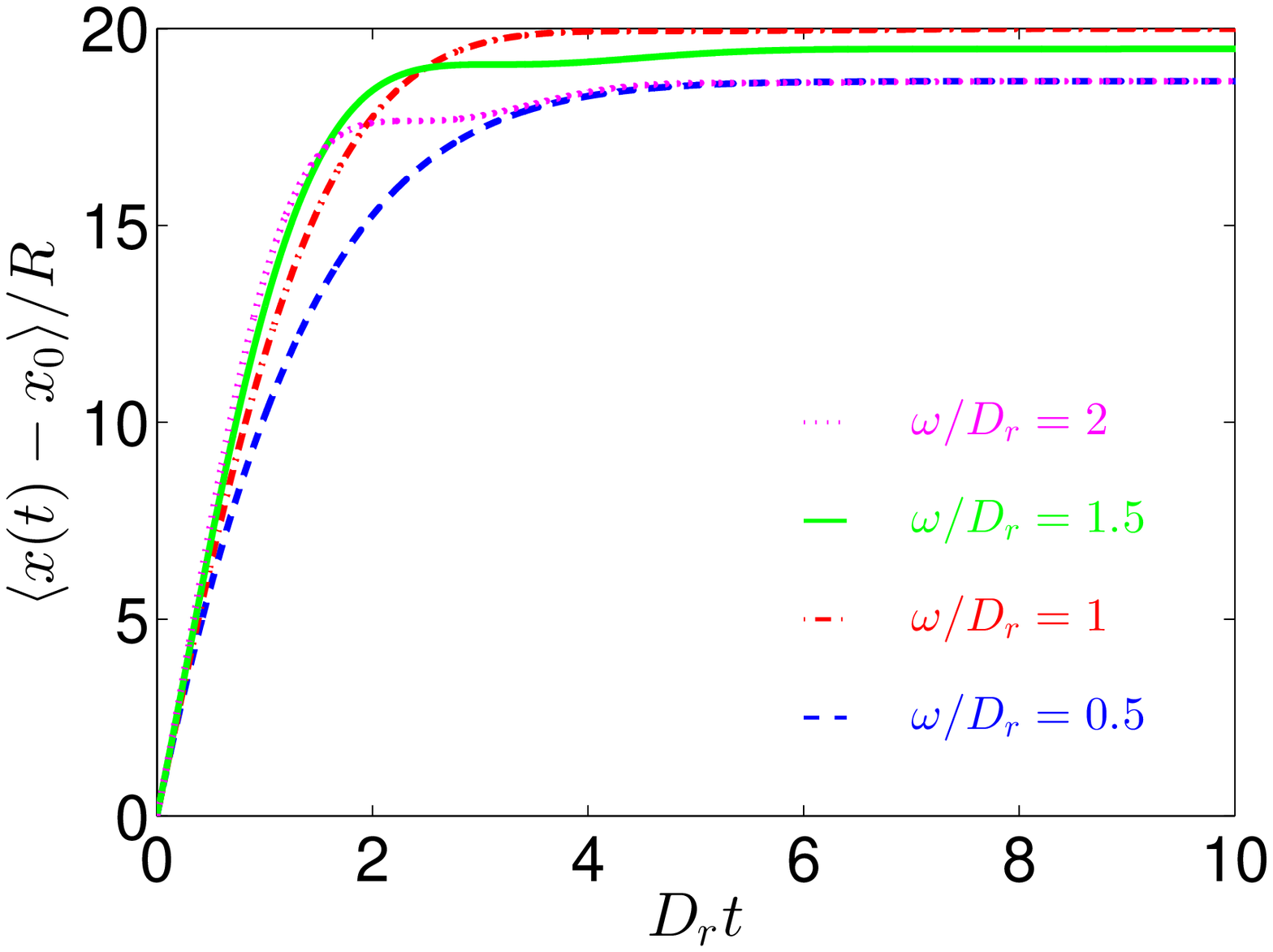}}
\caption{Mean position of a self-propelled particle with sinusoidal driving force for different values of the frequency $\omega$. The offset is $c=0$ in (a) and $c=1$ in (b) for fixed values of the scaled amplitude $\beta R F_0=10$ and the initial orientation $\phi_0=0$. Inset: characterization of the sinusoidal self-propulsion force.\label{fig4}}
\end{figure}
The analytical expression for the mean position (equation (\ref{eq:mpsin})) is visualized in figure \ref{fig4}. The curves initially increase and reach a constant final value for long times, after a transient regime where the effect of the specific periodic type of the self-propulsion is visible. 
The existence and the position of one or more local maxima for intermediate times depend on the value of $\omega$.
For large $\omega$ the first maximum occurs earlier in time and is more distinct. At short times, higher values of $\omega$ lead in general to higher values of the mean position than observed for smaller $\omega$. 
The final value for long times, which is analytically given by
\begin{eqnarray}
\lim\limits_{t\rightarrow\infty}\langle x(t)-x_0\rangle=\frac{4}{3}\beta F_0 R^2 D_\rmr  \cos(\phi_0)\left(\frac{\omega}{D_\rmr ^2+\omega^2}+\frac{c}{D_\rmr }\right)\,,
\end{eqnarray}
is maximal for $\omega=D_\rmr $. For $\omega = \kappa D_\rmr $ with an arbitrary value of $\kappa$ it is the same as for $\omega = D_\rmr /\kappa$.

\subsubsection{Mean square displacement}
\begin{figure}[tbh]
\centering
\subfigure[]{\includegraphics[width=0.45\textwidth]{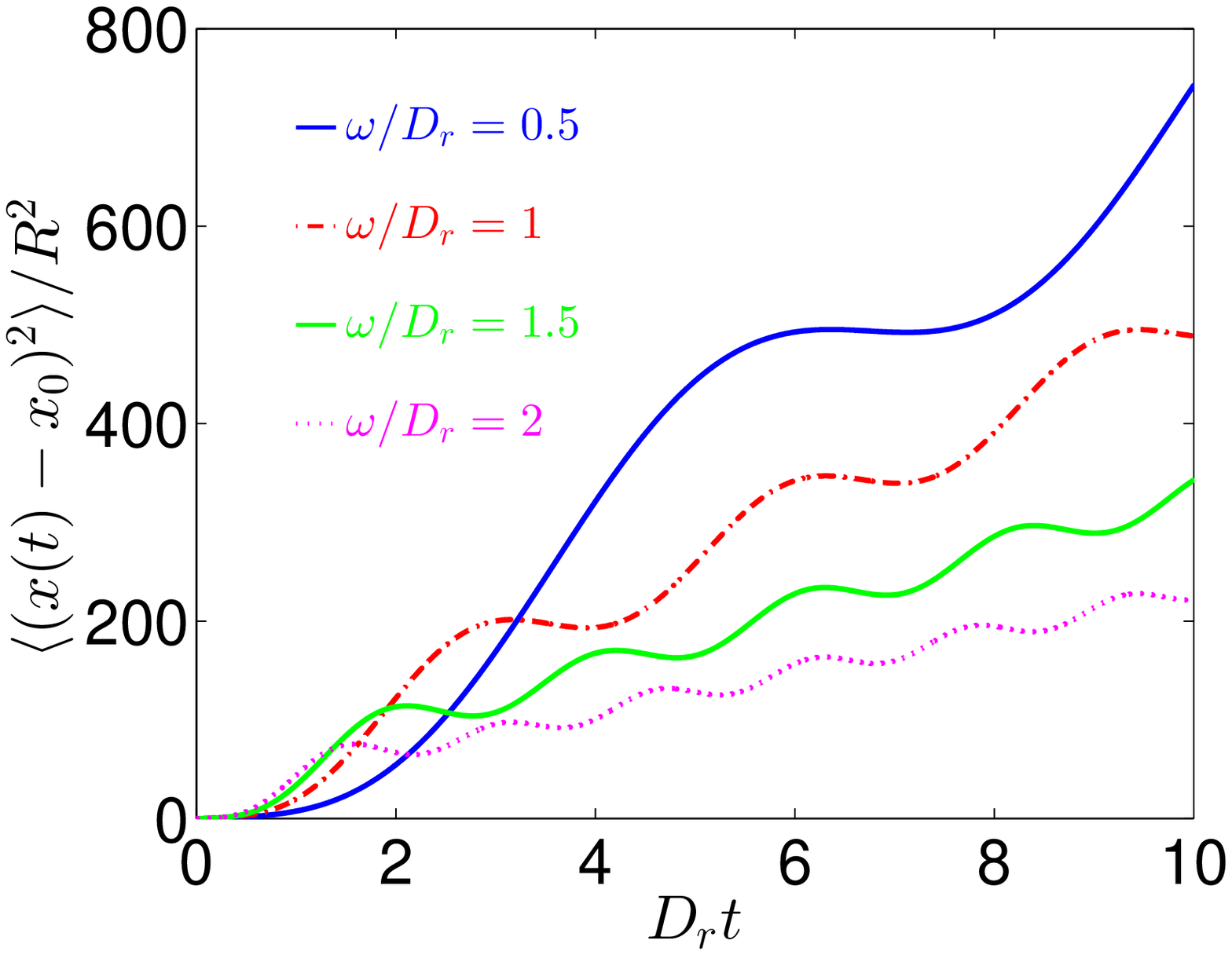}}
\qquad
\subfigure[]{\includegraphics[width=0.45\textwidth]{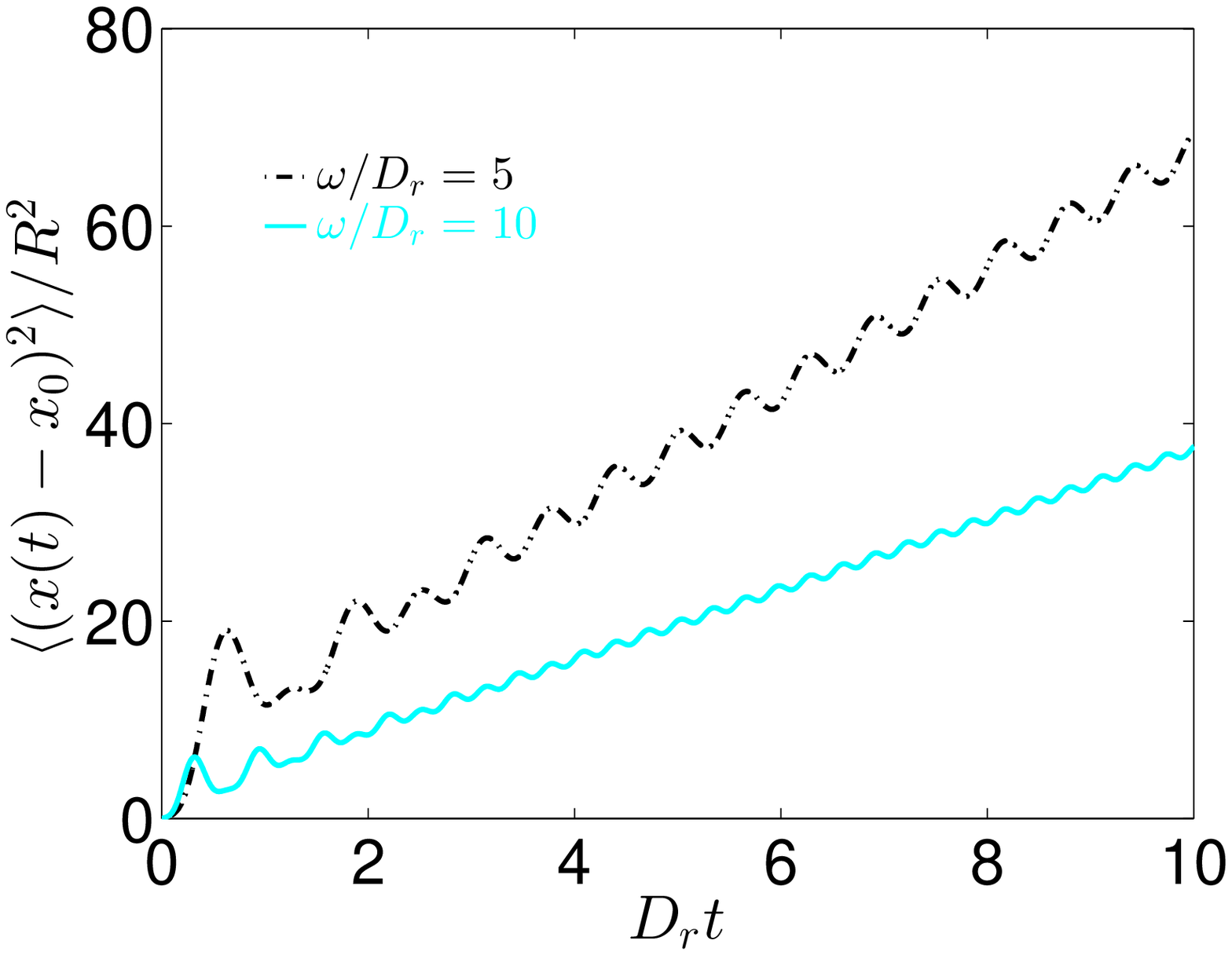}}
\caption{Mean square displacement of a self-propelled particle with sinusoidal driving force for different values of the frequency $\omega$. The parameters are $\beta R F_0=10$, $\phi_0=0$, and $c=0$. (a) Curves for low values of $\omega$ between $0.5$ and $2$. (b) For larger $\omega$, a transition from a slowly oscillating initial regime to a regular periodicity with double frequency is observed.\label{fig5}}
\end{figure}

\begin{figure}[tbh]
\centering
\includegraphics[width=0.45\textwidth]{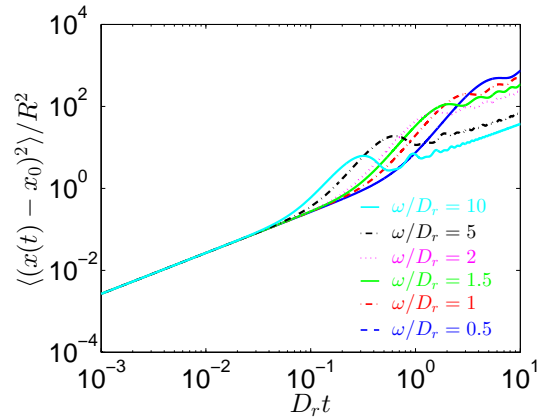}
\caption{Same curves as in figures $\ref{fig5}$(a) and $\ref{fig5}$(b), but now in a logarithmic representation.\label{fig6}}
\end{figure}

The mean square displacement of an active particle with sinusoidal self-propulsion force is
\begin{eqnarray}
\fl\langle(x(t)-x_0)^2\rangle=&2Dt +\frac{16}{9}\Big(\beta F_0R^2\Big)^2D_\rmr ^2\Bigg[\frac{1}{D_\rmr ^2+\omega^2}\left(\frac{D_\rmr t}{2}-\frac{D_\rmr \sin\left(2\omega t\right)}{4\omega}-\frac{1}{2}\sin^2\left(\omega t\right)\right)\nonumber\\
&-\frac{\omega}{D_\rmr ^2+\omega^2}\ \eta^-\left(1,1,1\right)+\frac{\cos\left(2\phi_0\right)}{9D_\rmr ^2+\omega^2}\Bigg(\frac{3}{2}\sin\left(\omega t\right)D_\rmr \ \eta^0\left(2,1,4\right)\nonumber\\
&+\frac{3}{2}\ \frac{\omega^2}{16D_\rmr ^2+4\omega^2}\left(\rme^{-4D_\rmr t}-1\right)+\frac{\omega}{4}\ \eta_2^-\left(2,1,4\right)-\omega\ \eta^-\left(1,1,1\right)\Bigg)\nonumber\\
&+\frac{c}{D_\rmr }\Bigg(\frac{1-\cos\left(\omega t\right)}{\omega}+\ \eta^-\left(1,1,1\right)\nonumber\\
&+\frac{1}{3}\cos\left(2\phi_0\right)\left(\  \eta^-\left(4,1,4\right)-\eta^-\left(1,1,1\right)\right)\Bigg)\nonumber\\
&+\frac{c}{D_\rmr ^2+\omega^2}\left(\frac{D_\rmr }{\omega}\left(1-\cos\left(\omega t\right)\right)-\sin\left(\omega t\right)+\frac{\omega}{D_\rmr }\left(1-\rme^{-D_\rmr t}\right)\right)\nonumber\\
&+\frac{c\ \cos\left(2\phi_0\right)}{\omega^2+9D_\rmr ^2}\left(3D_\rmr \ \eta^-\left(4,1,4\right)-\omega\ \tilde{\eta}\left(-4,1,4\right)\right)\nonumber\\
&+\frac{c\ \cos\left(2\phi_0\right)}{\omega^2+9D_\rmr ^2}\frac{\omega}{D_\rmr }\left(1-\rme^{-D_\rmr t}\right)+\frac{c^2 t}{D_\rmr }+\frac{c^2}{D_\rmr ^2}\left(\rme^{-D_\rmr t}-1\right)\nonumber\\
&+\frac{c^2\ \cos\left(2\phi_0\right)}{3D_\rmr ^2}\left(1-\rme^{-D_\rmr t}+\frac{1}{4}\left(\rme^{-4D_\rmr t}-1\right)\right)
\Bigg]
\label{eq:msdsin}
\end{eqnarray}
with the short notations
\begin{eqnarray}
\eta^-_a\left(b,c,d\right)&:=\frac{\left(bD_\rmr \sin\left(a\omega t\right)+c\omega\cos\left(a\omega t\right)\right)\rme^{-dD_\rmr t}-c\omega}{\left(c\omega^2\right)+\left(bD_\rmr \right)^2}\,,\\
\eta^0_a\left(b,c,d\right)&:=\frac{\left(bD_\rmr \sin\left(a\omega t\right)+c\omega\cos\left(a\omega t\right)\right)\rme^{-dD_\rmr t}}{\left(c\omega^2\right)+\left(bD_\rmr \right)^2}\,,\\
\tilde{\eta}_a(b,c,d)&:=\frac{\left(bD_\rmr \cos\left(a\omega t\right)+c\omega\sin\left(a\omega t\right)\right)\rme^{-dD_\rmr t}-bD_\rmr }{\left(c\omega\right)^2+\left(bD_\rmr \right)^2}\,,\\
\eta^-\left(b,c,d\right)&\equiv \eta^-_1\left(b,c,d\right)\,,\\
\eta^0\left(b,c,d\right)&\equiv \eta^0_1\left(b,c,d\right)\,,\\
\tilde{\eta}(b,c,d)&\equiv\tilde{\eta}_1(b,c,d)\,.
\end{eqnarray}
The result in equation (\ref{eq:msdsin}) is illustrated in figure \ref{fig5}. Obviously, for $\phi_0 = 0$ and $c=0$, smaller values of $\omega$ lead to a slower but longer initial increase due to the sine in equation (\ref{eq:Fsin}).
Consequently, for short times one obtains a larger mean square displacement for larger $\omega$ while for longer times more significant displacements result for smaller values of $\omega$ (see figure \ref{fig5}(a)). The curves for $\omega/D_\rmr  = 5$ and $\omega/D_\rmr  = 10$ in figure \ref{fig5}(b) yield some irregularities at short times: the first maximum of the oscillation is particularly large whereas the second one is much smaller than expected.
This behavior is induced by the sign changes of the sine in equation (\ref{eq:Fsin}). As the particle still has some memory of its initial orientation, the mean square displacement is significantly reduced when the propagation direction reverses. 
For long times, the rotational diffusion eliminates all orientation-dependent effects. Thus, a periodic behavior with double frequency occurs. It is no longer possible to distinguish between sign changes of the propulsion force from $+$ to $-$ and from $-$ to $+$.
The logarithmic plots in figure $\ref{fig6}$ represent the particularly large first oscillation even more clearly. It is most obvious for larger values of $\omega$ and is followed by a leveled second peak.

By neglecting the exponentially decreasing terms, the long-time behavior of the mean square displacement is obtained as 
\begin{eqnarray}
\fl\langle(x(t)-x_0)^2\rangle=&2Dt+\frac{16}{9}\Big(\beta F_0R^2\Big)^2D_\rmr ^2\Bigg[\frac{t}{2}\ \frac{D_\rmr }{D_\rmr ^2+\omega^2}+\frac{tc^2}{D_\rmr } + \textnormal{const.}\left(\omega,\phi_0,c\right)\nonumber\\
&-\sin\left(2\omega t\right)\left(\frac{D_\rmr }{4\omega}\ \frac{1}{D_\rmr ^2+\omega^2}\right)-\sin^2\left(\omega t\right)\left(\frac{1}{2}\frac{1}{D_\rmr ^2+\omega^2}\right)\nonumber\\
&-\sin\left(\omega t\right)\frac{c}{D_\rmr ^2+\omega^2}
-\cos\left(\omega t\right)\left(\frac{c}{D_\rmr \omega}+\frac{c\ D_\rmr }{\omega}\ \frac{1}{D_\rmr ^2+\omega^2}\right)\Bigg]\,.
\end{eqnarray}
\newline 
In the limit $\omega\rightarrow 0$ the solutions for a constant self-propulsion force are recovered.
Otherwise, for $\omega\ne 0$, the result for the long-time diffusion coefficient $D_\rml $ is
\begin{eqnarray}
D_\rml =\frac{1}{2t}\lim\limits_{t\rightarrow\infty}\langle(x(t) \! - \! x_0)^2\rangle=D \! + \! \Big(\beta F_0 R^2\Big)^2D_\rmr ^2\left(\frac{4}{9}\frac{D_\rmr }{D_\rmr ^2+\omega^2}+\frac{8}{9}\frac{c^2}{D_\rmr }\right),
\end{eqnarray} 
corresponding to a situation with a constant force $F= c F_0$ if $\omega\rightarrow\infty$. 

\subsection{Power-law self-propulsion force}
\label{power}
Finally, we consider a power-law time dependence
\begin{equation}
\label{eq:power}
F(t)=F_0\ (D_\rmr t)^\alpha\hspace{1cm}\textnormal{with } \alpha=0,1,2,\ldots
\end{equation}
and an arbitrary but constant prefactor $F_0$. 
In principle,  on large time scales the proportionality of the driving force to $t^\alpha$ with $\alpha > 0$ corresponds to a random walk with continuously increasing step size.

\subsubsection{Mean position}
\begin{figure}[tbh]\centering
\centering
\includegraphics[width=0.75\textwidth]{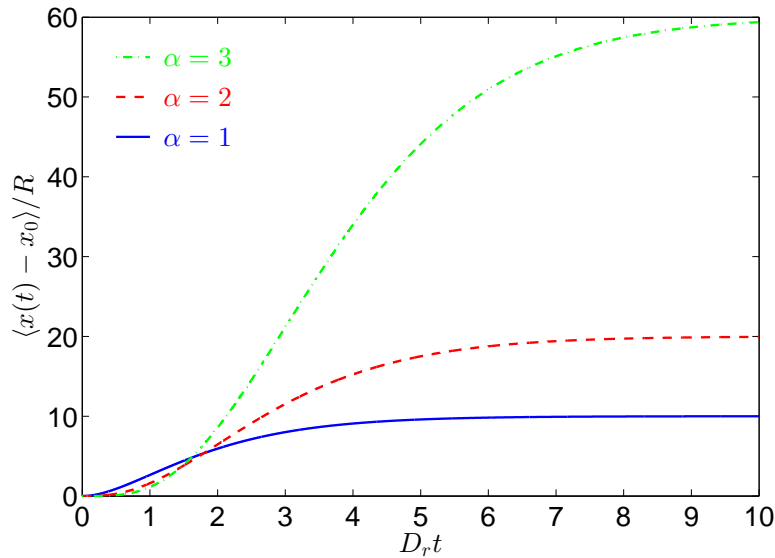}
\caption{Analytically calculated mean position of a self-propelled particle with a power-law self-propulsion force for different exponents $\alpha$. The constant parameters are $\beta R F_0=10$ and $\phi_0=0$.
 All curves approach a maximum value determined by equation (\ref{powerlim}).\label{fig:7}}
\end{figure}

Solving equations (\ref{1x})-(\ref{2}) for a self-propulsion according to equation (\ref{eq:power}) gives the mean position 
\begin{eqnarray}
\langle x(t)-x_0\rangle&=\frac{4}{3}\beta F_0 R^2 \cos \left(\phi_0\right)\left[\alpha!-\sum_{k=0}^{\alpha}(D_\rmr t)^{\alpha-k}\ \rme^{-D_\rmr t}\frac{\alpha!}{(\alpha-k)!}\right]
\end{eqnarray}
with the long-time limit
\begin{equation}
\lim\limits_{t\rightarrow \infty}\langle x(t)-x_0\rangle=\frac{4}{3}\beta F_0 R^2\cos\left(\phi_0\right)\alpha!\,.
\label{powerlim}
\end{equation}
For $\alpha=1,2,3$ the mean position is visualized in figure \ref{fig:7}. For short and intermediate times the curves increase until they reach a constant final value which depends on the specific exponent $\alpha$. The larger the value of $\alpha$ the longer lasts the initial stage and the higher is the final mean displacement. 

\subsubsection{Mean square displacement}
The one-dimensional mean square displacement for a power-law self-propulsion force is 
\begin{eqnarray}
\label{eq:1dMSD}
\fl\langle(x(t)-x_0)^2\rangle=&2Dt+\frac{16}{9}\left(\beta F_0 R^2\right)^2\Bigg\{\sum_{k=0}^{\alpha}(-1)^k\frac{\alpha!}{(\alpha-k)!}\frac{1}{2\alpha-k+1}(D_\rmr t)^{2\alpha-k+1}\nonumber\\
&+(-1)^\alpha \alpha!\Bigg(\sum_{k=0}^{\alpha}(D_\rmr t)^{\alpha-k}\rme^{-D_\rmr t} \frac{\alpha!}{(\alpha-k)!}-\alpha!\Bigg)\nonumber\\
&+\cos\left(2\phi_0\right)\Bigg[\sum_{k=0}^{\alpha}\frac{1}{3^{k+1}}\frac{\alpha!}{(\alpha-k)!}
\Bigg(\sum_{j=0}^{2\alpha-k}\frac{1}{4}(D_\rmr t)^{2\alpha-k-j}\rme^{-4D_\rmr t}\frac{(2\alpha-k)!}{(2\alpha-k-j)!}\nonumber\\
&-\frac{1}{4}(2\alpha-k)!\Bigg) -\frac{\alpha!}{3^{\alpha+1}}\Bigg(\sum_{k=0}^{\alpha}(D_\rmr t)^{\alpha-k}\rme^{-D_\rmr t}\frac{\alpha!}{(\alpha-k)!}-\alpha!\Bigg)\Bigg]\Bigg\}\,.
\end{eqnarray}
While equation (\ref{eq:1dMSD}) depends on the initial orientation $\phi_0$ of the particle, the mean square displacement can also be given in the two-dimensional version 
\begin{eqnarray}
\label{eq:2dMSD}
\fl\langle(\mathbf{r}(t)-\mathbf{r}_0)^2\rangle=&4Dt+\frac{32}{9}\left(\beta F_0 R^2\right)^2\Bigg[\sum_{k=0}^{\alpha}(-1)^k\frac{\alpha!}{(\alpha-k)!\,(2\alpha-k+1)}(D_\rmr t)^{2\alpha-k+1}\nonumber\\
&+(-1)^\alpha\alpha!\Bigg(\sum_{k=0}^{\alpha}(D_\rmr t)^{\alpha-k}\rme^{-D_\rmr t} \frac{\alpha!}{(\alpha-k)!}-\alpha!\Bigg)\Bigg]\,,
\end{eqnarray}
\begin{figure}[bt]
\centering
\includegraphics[width=0.75\textwidth]{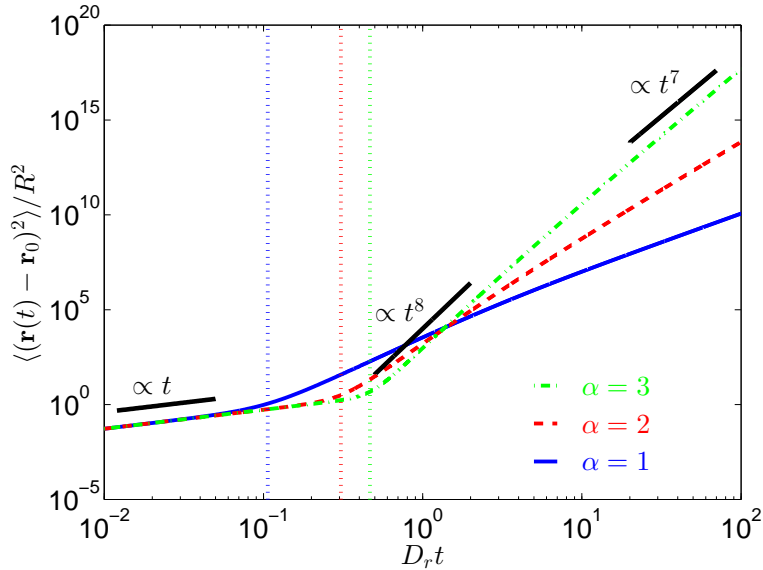}
\caption{Mean square displacement of a self-propelled particle with a power-law driving force in a logarithmic representation. The strength of the self-propulsion is given by $\beta R F_0=100$. Three different regimes $\propto \! t$, $\propto \! t^{2\alpha+2}$, and $\propto \! t^{2\alpha+1}$ are identified and explicitly indicated for the exponent $\alpha=3$.\label{fig8}}
\end{figure}
which is independent of the initial conditions. 
The visualization of equation (\ref{eq:2dMSD}) in figure \ref{fig8} exhibits three qualitatively different time regimes. A diffusive regime at short times is followed by a superdiffusive $\propto \!\! t^{2\alpha+2}$ regime. Finally, for $t>1/D_\rmr $, which corresponds to the characteristic time scale for the rotational Brownian motion, the curves enter another superdiffusive regime, where the scaling is $\propto \! t^{2\alpha+1}$. Whereas the time for this last transition does not depend on the exponent $\alpha$ of the self-propulsion force, the crossover from the diffusive to the first superdiffusive regime occurs at the time 
\begin{eqnarray}
\label{eq:trans}
D_\rmr t^*(\alpha)=\left[\frac{2}{3} (\beta F_0 R)^2 \left(\sum_{k=0}^{\alpha}(-1)^k\frac{(\alpha!)^2}{(\alpha-k)!(\alpha+k+2)!}\right)\right]^{-1/(2\alpha+1)}\,,
\end{eqnarray}
which is determined by the specific type of power law and the propulsion strength. 
For the various curves in figure \ref{fig8} the transition time according to equation (\ref{eq:trans}) is indicated by vertical lines.

\section{Results for an additional constant torque}\label{time-dep+torque}
\label{sec:torque}
For many experimental systems it is possible to describe the motion of the respective natural or artificial microswimmers by implementing only an effective self-propulsion force corresponding to a translational swimming velocity \cite{Howse_2007,Golestanian_2012,Silber-Li}  
in the Langevin equations. However, a more detailed investigation often yields that either particle imperfections or asymmetric shapes \cite{Kuemmel:13,Kraft:13,Wagner2013,Chakrabarty2013} induce a deterministic rotational motion of the swimming object.
To account for this, an additional torque $\mathbf{M}=M\hat{\mathbf{e}}_z$ (see figure \ref{fig:model}) has to be considered in the orientational Langevin equation, while 
equations (\ref{1x}) and (\ref{1y}) stay the same. The updated version of equation (\ref{2}) is given by 
\begin{eqnarray}
\frac{d\phi}{dt}=\beta D_\rmr \left[ M+g(t)\right]\,, \label{2+t}
\end{eqnarray}
which leads to $\langle\phi(t)\rangle=\phi_0+ \beta D_\rmr  M t$ and $\langle(\phi(t)-\langle\phi(t)\rangle)^2\rangle=2D_\rmr t$
for the first and second moments of the angular displacement distribution.   
As a constant torque does not destroy the Gaussianity of the orientational distribution \cite{tenHagen_JPCM}, the Langevin equations can be solved similarly to the torque-free case discussed in section \ref{time-dep}. In the following, this is done exemplarily for the sinusoidal self-propulsion force (see section \ref{sinusoidal}).

\subsection{Trajectories for vanishing noise}
To gain a better understanding of the interplay between the oscillating driving force and an additional constant torque, we first consider the noise-free case by neglecting the random terms in equations (\ref{1x}), (\ref{1y}), and (\ref{2+t}). With $\nu = \beta D_\rmr  M$ this leads to  
\begin{equation}
\phi(t)=\phi_0+\nu t
\end{equation}
for the rotational motion and 
\begin{eqnarray}
\label{eq:mpnoisefree}
\fl x(t)-x_0&=\beta D F_0\Bigg\{\cos\left(\phi_0\right)\left[-\frac{1}{2}\left(\frac{\cos(t(\omega-\nu))}{\omega-\nu}+\frac{\cos(t(\omega+\nu))}{\omega+\nu}\right)+\frac{\omega}{\omega^2-\nu^2}\right]\nonumber\\
\fl
& \quad -\sin\left(\phi_0\right)\frac{1}{2}\left(\frac{\sin\left(t(\omega-\nu)\right)}{\omega-\nu}-\frac{\sin\left(t(\omega+\nu)\right)}{\omega+\nu}\right)\nonumber\\
\fl
& \quad +\frac{c}{\nu}\left(\sin\left(\phi_0+\nu t\right)-\sin\left(\phi_0\right)\right)\Bigg\}
\end{eqnarray}
for the translational particle displacement.

\begin{figure}[tbh]
\centering
\subfigure[]{\includegraphics[width=0.45\textwidth]{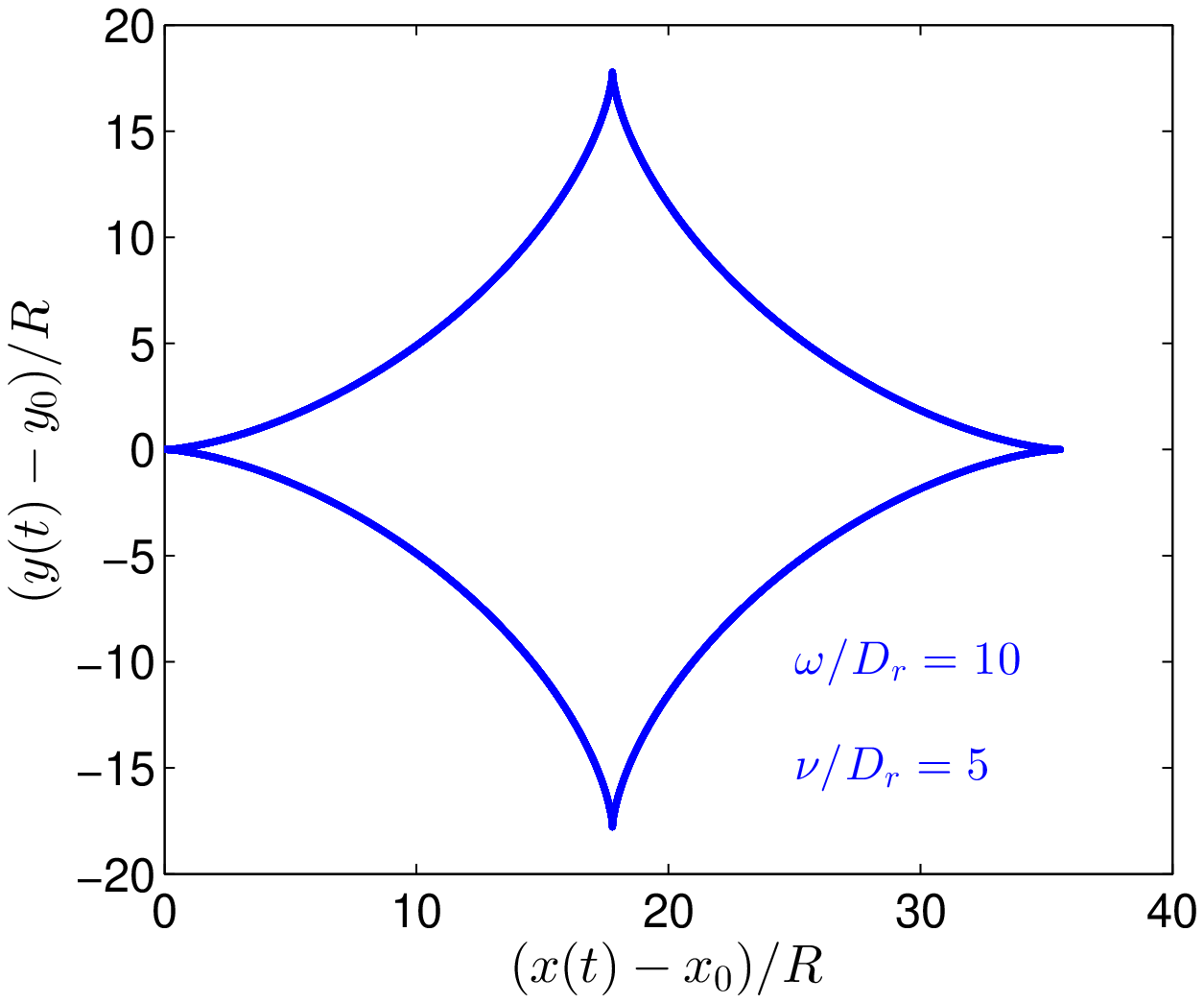}}
\qquad
\subfigure[]{\includegraphics[width=0.45\textwidth]{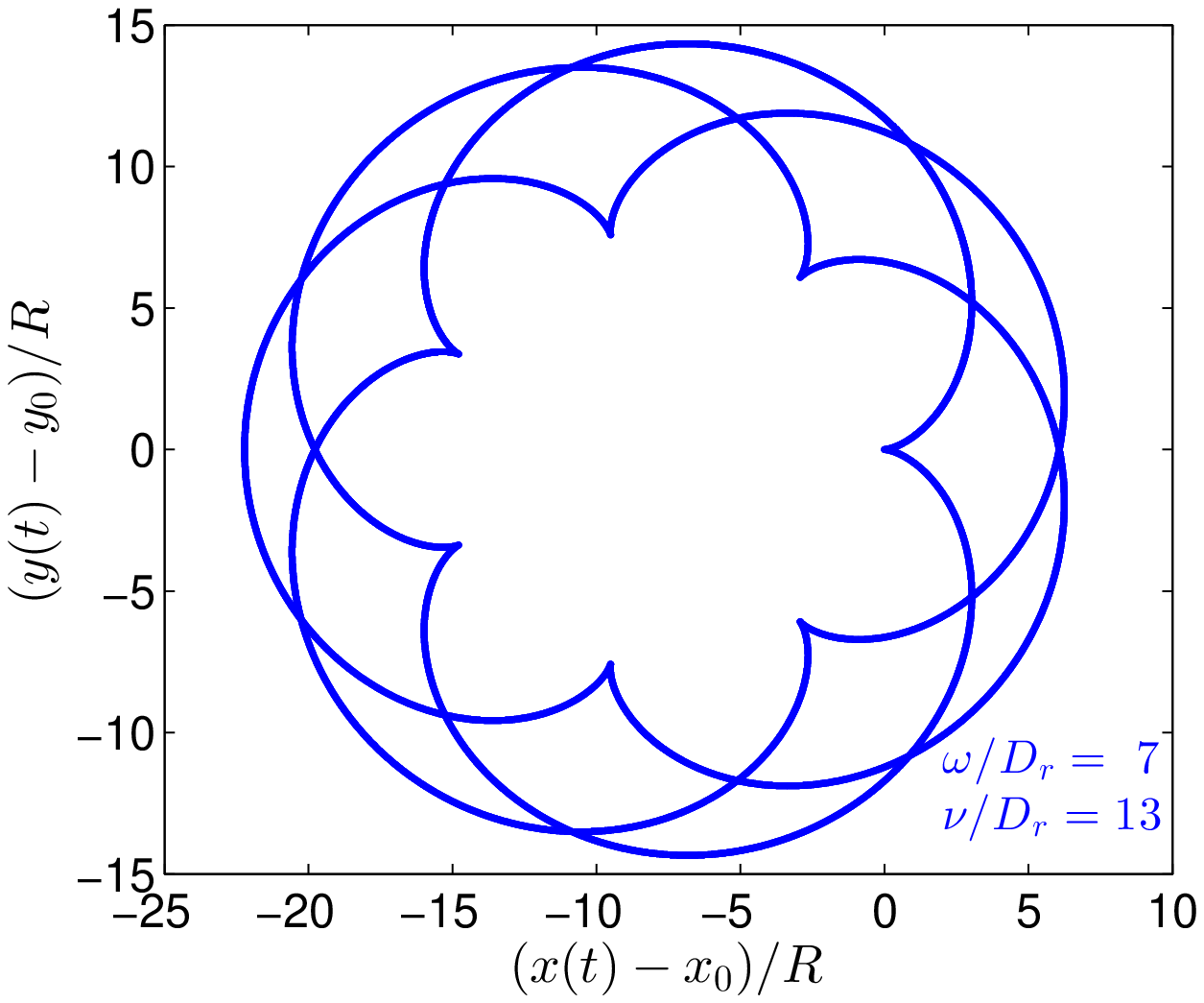}}
\caption{Noise-free trajectories of a self-propelled particle with sinusoidal self-propulsion and an additional constant torque. The plots are based on equation (\ref{eq:mpnoisefree}). Curves are shown for $\beta R F_0=100$, $\phi_0=0$, $c=0$, and different values of $\omega$ and $\nu$:  (a) $\omega=10D_\rmr $ and $\nu=5D_\rmr $; (b) $\omega=7D_\rmr $ and $\nu=13D_\rmr $. For the case $c=0$, closed trajectories are obtained as long as $\omega\ne\nu$.\label{fig:9}}
\end{figure}

\begin{figure}[tbh]\centering
\centering
\includegraphics[width=0.45\textwidth]{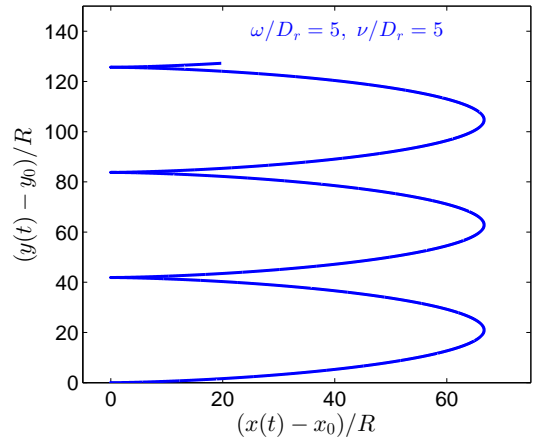}
\caption{Same as in figure \ref{fig:9}, but now for $\omega=\nu=5D_\rmr $. The special case $\omega=\nu$ is the only situation for which the trajectories are not closed.\label{fig:10}}
\end{figure}

As illustrated in figure \ref{fig:9}, for $c=0$ the particle moves on closed trajectories which display a certain number of vertices in a regular pattern. They occur whenever the sign of the propulsion force changes, i.e., in time steps of $\pi/\omega$. 
If $\omega>\nu$, the vertices face outward; for $\omega<\nu$ they face inward.
The number $N$ of vertices per closed loop depends on the frequencies $\omega$ and $\nu$ according to
\begin{eqnarray}
\label{eq:edgenumber}
N&=2\omega\ \textnormal{lcm}\left(\frac{1}{\omega+\nu},\frac{1}{|\omega-\nu|}\right)\nonumber\\
&=\frac{2\omega}{(\omega+\nu)|\omega-\nu|}\ \textnormal{lcm}\left(|\omega-\nu|, \omega+\nu\right)\,.
\end{eqnarray}
Here, lcm denotes the least common multiple as the product of the highest order of each prime factor. It is generalized to fractions by also allowing negative exponents.

Closed trajectories are always obtained as long as $\omega \ne \nu$. However, for the special case $\omega = \nu$ the propulsion direction reverses just at the moment when the orientation has changed by $\unit{180}{\degree}$. Thus, a trajectory with a continuously increasing displacement in one direction is established (see figure \ref{fig:10}).

\subsection{Mean position}
\begin{figure}[tbh]
\centering
\subfigure[]{\includegraphics[width=0.85\textwidth]{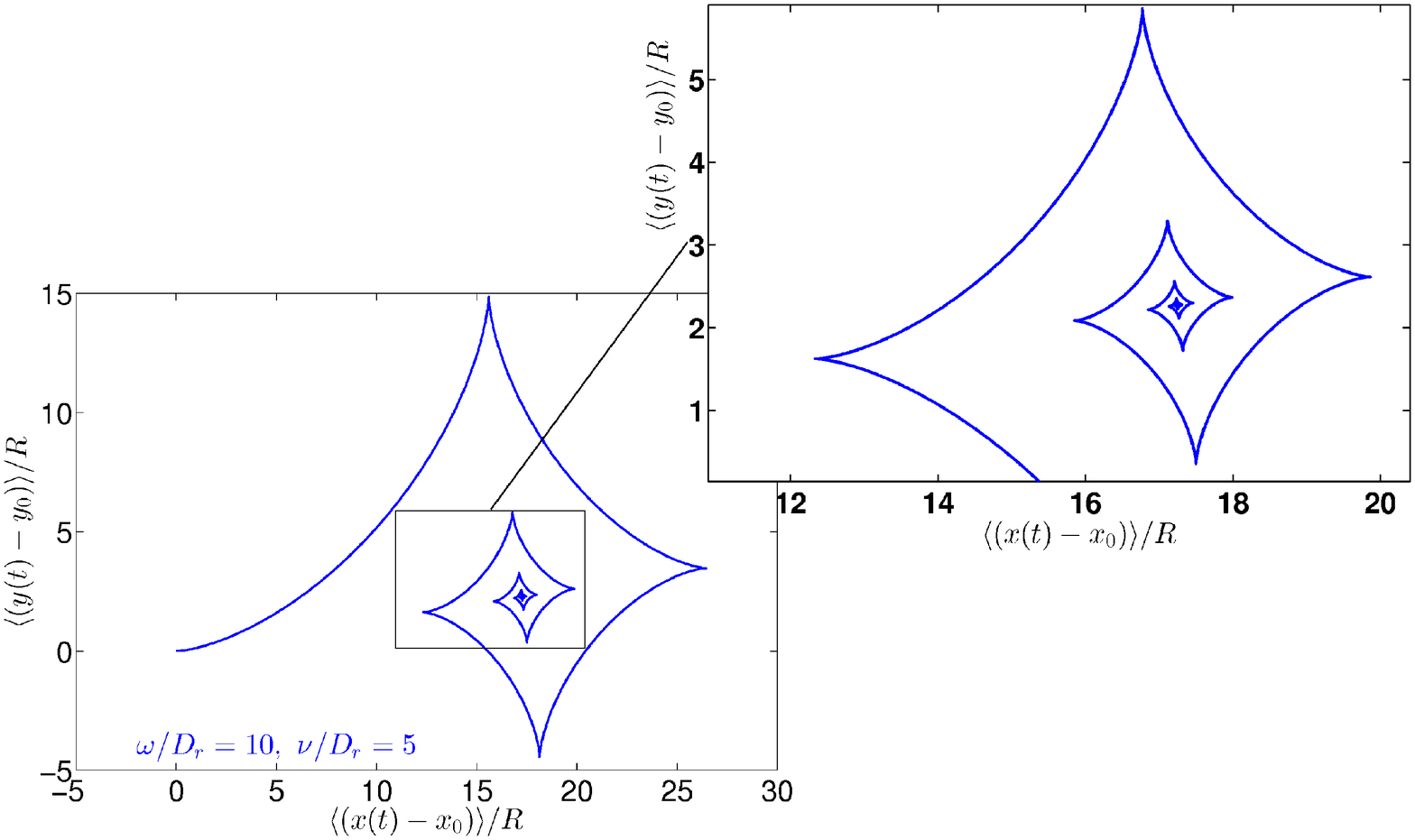}}
\subfigure[]{\includegraphics[width=0.85\textwidth]{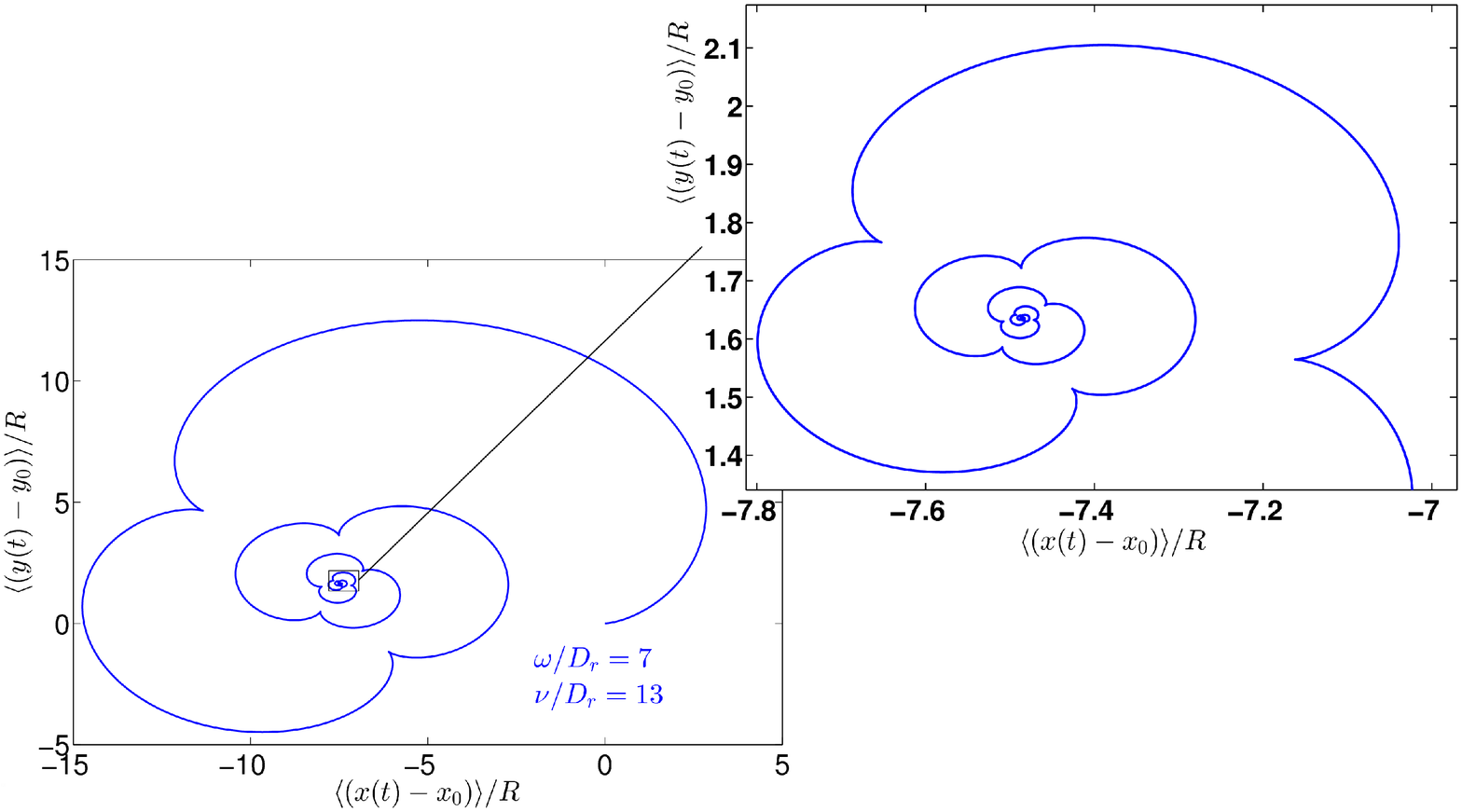}}
\caption{Mean trajectories of a self-propelled particle with sinusoidal self-propulsion and an additional constant torque for $\beta R F_0=100$, $\phi_0=0$, $c=0$, and different values of $\omega$ and $\nu$: (a) $\omega=10D_\rmr $ and $\nu=5D_\rmr $; (b) $\omega=7D_\rmr $ and $\nu=13D_\rmr $. The curves are self-similar, as illustrated by the closeups of the framed regions, and bear the same characteristics as their noise-free counterparts (see figure \ref{fig:9}). 
\label{fig:11}}
\end{figure}
\begin{figure}[tbh]\centering
\centering
\includegraphics[width=0.85\textwidth]{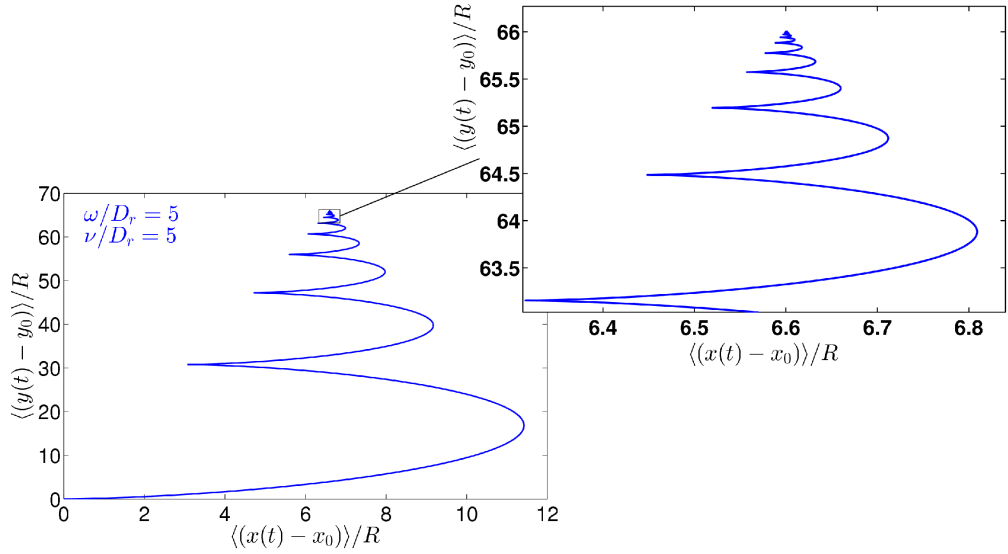}
\caption{Noise-averaged trajectory of a self-propelled particle with sinusoidal self-propulsion and an additional constant torque for $\omega=\nu=5D_\rmr $, $\beta R F_0=100$, $\phi_0=0$, and $c=0$. Similar to the noise-free counterpart (see figure \ref{fig:10}), one obtains a trajectory with a preferred direction of translation. The curve is self-similar, as visualized in the closeup of the framed area in the plot.\label{fig:12}}
\end{figure}

For non-zero noise, the mean position of an active particle with self-propulsion as defined in equation (\ref{eq:Fsin}) is a linear superposition of the contributions originating from a purely sinusoidal force $F(t) = F_0 \sin (\omega t)$ on the one hand and a constant force $F=cF_0$ on the other hand. As the latter case has already been considered in reference \cite{Cond_Matt}, here we present only the result for $c=0$:  
\begin{eqnarray}
\label{eq:mpselfsim}
\fl\langle x(t)-x_0\rangle&=\beta D F_0\Bigg[\frac{\rme^{-D_\rmr t}}{2} \Bigg(-\frac{D_\rmr  \sin\left(t(\omega-\nu)-\phi_0\right)+(\omega-\nu)\cos\left(t(\omega-\nu)-\phi_0\right)}{(\omega-\nu)^2+D_\rmr ^2}\nonumber\\
\fl
&\quad -\frac{D_\rmr\sin\left(t(\omega+\nu)+\phi_0\right)+(\omega+\nu)\cos\left(t(\omega+\nu)+\phi_0\right)}{(\omega+\nu)^2+D_\rmr ^2}\Bigg)\nonumber\\
\fl
&\quad +\frac{1}{2}\frac{(\omega-\nu)\cos\left(\phi_0\right)-D_\rmr\sin(\phi_0)}{(\omega-\nu)^2+D_\rmr ^2}+\frac{1}{2}\frac{(\omega+\nu)\cos\left(\phi_0\right)+D_\rmr\sin(\phi_0)}{(\omega+\nu)^2+D_\rmr ^2}\Bigg]\,.
\end{eqnarray} 
The corresponding mean trajectories are similar to the noise-free ones (see figures \ref{fig:11} and \ref{fig:12} as compared to figures \ref{fig:9} and \ref{fig:10}, respectively). However, when taking the Brownian random terms into account, we do not obtain closed mean swimming paths. Instead of that, the size of the curves reduces exponentially. 
As can be seen clearly in figures \ref{fig:11} and \ref{fig:12}, the mean trajectories are \textit{self-similar}. This characteristic feature also follows directly from the analytical expression in equation (\ref{eq:mpselfsim}). The scaling factor for the self-similarity is $\rme^{-D_\rmr t}$. All other terms are periodic in time $t$ with a period of either $T_1=2\pi /(\omega + \nu)$ or $T_2=2\pi /|\omega - \nu|$. Thus, after
\begin{equation}
T=2\pi\ \textnormal{lcm}\left(\frac{1}{\omega+\nu},\frac{1}{|\omega-\nu|}\right)    
\end{equation}
the scaled trajectory overlaps with itself. 

The self-similarity is an important property of mean microswimmer trajectories, also in the context of a comparison with the situation of a constant force, where the mean swimming path was shown to be a logarithmic spiral \cite{vanTeeffelenL2008}. 
While the latter is one of the simplest realizations of a self-similar curve, it is not intuitive that this feature also survives when a sinusoidal self-propulsion force is considered.

\subsection{Mean square displacement}
\begin{figure}[tbh]
\centering
\includegraphics[width=0.75\textwidth]{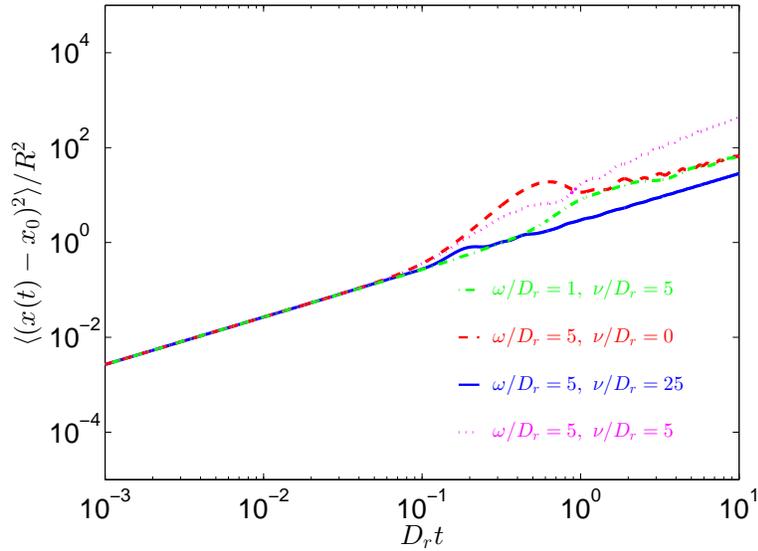}
\caption{Mean square displacement of a self-propelled particle with sinusoidal self-propulsion force and an additional constant torque for different values of $\omega$ and $\nu$. The fixed parameters are given by $\beta R F_0=10$, $\phi_0=0$, and $c=0$. At long times, the mean square displacement is much larger for the special case $\omega=\nu$ than for $\omega\ne\nu$.\label{fig:13}}
\end{figure}
The analytical expression for the mean square displacement of a self-propelled particle with sinusoidal self-propulsion and an additional constant torque is given in equation (\ref{eq:MSDaddtorque}) in the appendix.
Figure \ref{fig:13} shows the corresponding curves for different values of the frequencies $\omega$ and $\nu$. 
At long times, the special case $\omega=\nu$ induces much larger values for the mean square displacement than obtained for $\omega\ne\nu$. This can easily be explained by comparing the mean trajectories in figure \ref{fig:11} with figure \ref{fig:12}. Only the situation $\omega=\nu$, where the sign of the self-propulsion changes exactly after half a revolution of the particle, generates a motion primarily in one specific direction. This results in a much higher value for the mean square displacement. 

The long-term diffusion coefficient is analytically given by
\begin{eqnarray}
D_\rml =D+\frac{D_\rmr }{8}\beta^2D^2F_0^2\left[\frac{1}{(\omega-\nu)^2+D_\rmr ^2}+\frac{1}{(\omega+\nu)^2+D_\rmr ^2}+\frac{4c^2}{D_\rmr ^2+\nu^2}\right]\,.
\label{eq:Dcons}
\end{eqnarray}
For $\omega=\nu$ the first term inside the square brackets in equation (\ref{eq:Dcons}) becomes maximal because only the squared rotational diffusion coefficient remains in the denominator.
In contrast, a strong torque significantly reduces the mean square displacement, as illustrated by the solid curve in figure \ref{fig:13}. 

\section{Conclusion}
\label{summary}
In conclusion, we have studied the influence of a time-dependent self-propulsion on the Brownian dynamics of an active colloidal particle. Our model based on the coupled translational and rotational Langevin equations provides analytical solutions for the mean position and the mean square displacement of swimmers with either time-periodic or continuously increasing propagation speed. Thus, previous coarse-grained models with constant self-propulsion are generalized towards a more realistic description of the propulsion mechanism, also on the time scale of an individual swimming stroke. 
The analysis yields that the noise-free path of a time-dependent swimmer can be quite complex, involving trajectories much more complicated than the commonly observed straight lines and circles. 
Moreover, we have analytically calculated the noise-averaged trajectories for time-periodic propulsion under the action of an additional torque. Interestingly, such an oscillatory microswimmer moves on average on a self-similar curve. 
If the effective self-propulsion force scales in time as $\propto \! t^\alpha$, superdiffusive behavior is found in the long-time regime where the mean square displacement reveals a $\propto \!\! t^{2\alpha +1}$ time dependence after an intermediate regime with a scaling $\propto \! t^{2\alpha +2}$. These new exponents are expected to also affect the non-Gaussian behavior of self-propelled particles \cite{Cond_Matt,Silber-Li}. 

An interesting next step would be to include a time dependence not only with regard to the self-propulsion force but also for the additional torque. Such a variation in time can either be externally prescribed by a magnetic field \cite{Baraban_ACSnano,Baraban_2013}, for example, or it can be of stochastic nature, as observed for slightly curved rods which undergo fluctuation-induced flipping leading to two equivalent stable states with an opposite sign of the torque \cite{TakagiPRL}. 

\section*{Acknowledgment}
This work was supported by the ERC Advanced Grant INTERCOCOS (Grant No.\ 267499). 

\appendix
\section*{Appendix}
\setcounter{section}{1}
Here, we present the analytical result for the mean square displacement of a self-propelled particle with sinusoidal self-propulsion under the action of an additional constant torque. It is given by 
\begin{eqnarray}
\fl\langle (x(t)-x_0)^2\rangle=&2Dt+\beta^2D^2F_0^2\Bigg\{\frac{D_\rmr }{2}\ \omega_1^+\left(  
\frac{t}{2}-\frac{\sin(2\omega t)}{4\omega}+\frac{c}{\omega} 
\left(1-\cos\left(\omega t\right)\right)\right)\nonumber\\
&-\frac{1}{2}\tilde{\omega}_1^+\left(\frac{\sin^2\left(\omega t\right)}{2\omega}+\frac{c}{\omega}\sin\left(\omega t\right)\right)+\frac{c\ D_\rmr }{D_\rmr ^2+\nu^2}\left(\frac{1}{\omega}\left(1-\cos\left(\omega t\right)\right)+ct\right)\nonumber\\
&+\left(\frac{1}{2}\tilde{\omega}_1^+-\frac{c\ D_\rmr }{D_\rmr ^2+\nu^2}\right)\left(A_1+cB_1\right)
+\left(\frac{1}{2}D_\rmr \ \omega^-_1+\frac{c\ \nu}{D_\rmr ^2+\nu^2}\right)\left(A_2+cB_2\right)\nonumber\\
&+\frac{1}{2}\sum\limits_{k=\pm1}\frac{1}{\left(\omega-k\nu\right)^2+9D_\rmr ^2}\Big[-3D_\rmr \Big(Z_2(4D_\rmr ,-2\phi_0,\omega-2k\nu)\nonumber\\
&+c\ Y_1(-4D_\rmr ,-2\phi_0,\omega-2k\nu)\Big)\nonumber\\
&-\left(\omega-k\nu\right)\Big(Z_1(4D_\rmr ,-2\phi_0,\omega-2k\nu)
+cY_2(-4D_\rmr ,-2\phi_0,\omega-2k\nu)\Big)\Big]\nonumber\\
&+\frac{1}{2}\tilde{\omega}_3^+\Big(Z_1(D_\rmr ,2\phi_0,\nu)+c\ Y_2(-D_\rmr ,2\phi_0,\nu)\Big)\nonumber\\
&+\frac{3}{2}D_\rmr \ \omega^-_3\Big(Z_2(D_\rmr ,2\phi_0,\nu)
+c\ Y_1(-D_\rmr ,2\phi_0,\nu)\Big)\nonumber\\
&+\frac{c}{9D_\rmr ^2+\nu^2}\Big[-3D_\rmr \Big(Z_1(4D_\rmr ,2\phi_0,2\nu)+c\ Y_2(-4D_\rmr ,2\phi_0,2\nu)\nonumber\\
&-Z_1(D_\rmr ,2\phi_0,\nu)-c\ Y_2(-D_\rmr ,2\phi_0,\nu)\Big)\nonumber\\
&+\nu\ \Big(Z_2(4D_\rmr ,2\phi_0,2\nu)+c\ Y_1(-4D_\rmr ,2\phi_0,2\nu)\nonumber\\
&-Z_2(D_\rmr ,2\phi_0,\nu)-c\ Y_1(-D_\rmr ,2\phi_0,\nu)\Big)\Big]\Bigg\} \,, \label{eq:MSDaddtorque}
\end{eqnarray}
where
\begin{eqnarray}
\omega_a^+ &=\frac{1}{(\omega+\nu)^2+(aD_\rmr)^2}+\frac{1}{(\omega-\nu)^2+(aD_\rmr)^2}\,,\\
\tilde{\omega}_a^+ &= \frac{\omega+\nu}{(\omega+\nu)^2+(aD_\rmr)^2}+\frac{\omega-\nu}{(\omega-\nu)^2+(aD_\rmr)^2}\,,\\
\omega_a^-&=\frac{1}{(\omega+\nu)^2+(aD_\rmr) ^2}-\frac{1}{(\omega-\nu)^2+(aD_\rmr) ^2}\,,\\
A_1&=\frac{\rme^{-D_\rmr t}}{2}\Bigg[-
\frac{D_\rmr \sin\left(t\left(\omega-\nu\right)\right)}{(\omega-\nu)^2+D_\rmr ^2}-
\frac{\left(\omega-\nu\right)\cos\left(t(\omega-\nu)\right)}
{\left(\omega-\nu\right)^2+D_\rmr ^2}\nonumber\\
& \quad -\frac{D_\rmr \sin\left(t\left(\omega+\nu\right)\right)}{(\omega+\nu)^2+D_\rmr ^2}-
\frac{\left(\omega+\nu\right)\cos\left(t(\omega+\nu)\right)}
{\left(\omega+\nu\right)^2+D_\rmr ^2}\Bigg]\nonumber\\
& \quad +\frac{1}{2}\frac{\omega-\nu}{\left(\omega-\nu\right)^2+D_\rmr ^2}
+\frac{1}{2}\frac{\omega+\nu}{\left(\omega+\nu\right)^2+D_\rmr ^2}\,,\\
A_2 & =\frac{\rme^{-D_\rmr t}}{2}\Bigg[
\frac{\left(\omega-\nu\right)\sin\left(t\left(\omega-\nu\right)\right)}{(\omega-\nu)^2+D_\rmr ^2}-
\frac{D_\rmr \cos\left(t(\omega-\nu)\right)}
{\left(\omega-\nu\right)^2+D_\rmr ^2}\nonumber\\
& \quad -\frac{\left(\omega+\nu\right)\sin\left(t\left(\omega+\nu\right)\right)}{(\omega+\nu)^2+D_\rmr ^2}+
\frac{D_\rmr \cos\left(t(\omega+\nu)\right)}
{\left(\omega+\nu\right)^2+D_\rmr ^2}\Bigg]\nonumber\\
& \quad +\frac{1}{2}\frac{D_\rmr }{\left(\omega-\nu\right)^2+D_\rmr ^2}
-\frac{1}{2}\frac{D_\rmr }{\left(\omega+\nu\right)^2+D_\rmr ^2}\,,\\
B_1&=\frac{\rme^{-D_\rmr  t}}{D_\rmr ^2+\nu^2}\Big(\nu\sin\left(\nu t\right)-D_\rmr \cos\left(\nu t\right)\Big)+\frac{D_\rmr }{D_\rmr ^2+\nu^2}\,,\\
B_2&=\frac{\rme^{-D_\rmr  t}}{D_\rmr ^2+\nu^2}\Big(-D_\rmr \sin\left(\nu t\right)-\nu\cos\left(\nu t\right)\Big)+\frac{\nu}{D_\rmr ^2+\nu^2}\,,
\end{eqnarray}
\begin{eqnarray}
\fl
Y_1(a,b,c) & = \frac{\rme^{at}}{a^2+c^2}\left(a\ \sin(ct+b)-c\cos(ct+b)\right)
-\frac{a\ \sin(b)-c\cos(b)}{a^2+c^2}\,, \\
\fl
Y_2(a,b,c) & = \frac{\rme^{at}}{a^2+c^2}\left(a\ \cos(ct+b)+c\sin(ct+b)\right)
-\frac{a\ \cos(b)+c\sin(b)}{a^2+c^2}\,, \\
\fl
Z_1(a,b,c) & = \frac{\rme^{-at}}{2}\Bigg[\frac{-a\sin\left(t(\omega-c)-b\right)}
{(\omega-c)^2+a^2}-\frac{\left(\omega-c\right)\cos\left(t(\omega-c)-b\right)}{\left(\omega-c\right)^2+a^2}\nonumber\\
\fl
&\quad -\frac{a\sin\left(t(\omega+c)+b\right)}
{(\omega+c)^2+a^2}-\frac{\left(\omega+c\right)\cos\left(t(\omega+c)+b\right)
}{\left(\omega+c\right)^2+a^2}\Bigg]\nonumber\\
\fl
&\quad+\frac{1}{2}\frac{\left(\omega-c\right)\cos\left(b\right)}{(\omega-c)^2+a^2}
+\frac{1}{2}\frac{\left(\omega+c\right)\cos\left(b\right)}{(\omega+c)^2+a^2}\nonumber\\
\fl
&\quad -\frac{1}{2}\frac{a\sin\left(b\right)}{(\omega-c)^2+a^2}
+\frac{1}{2}\frac{a\sin\left(b\right)}{(\omega+c)^2+a^2}\,, \\
\fl
Z_2(a,b,c) & = \frac{\rme^{-at}}{2}\Bigg[\frac{-a\cos\left(t(\omega-c)-b\right)}
{(\omega-c)^2+a^2}+\frac{\left(\omega-c\right)\sin\left(t(\omega-c)-b\right)}{\left(\omega-c\right)^2+a^2}\nonumber\\
\fl
&\quad +\frac{a\cos\left(t(\omega+c)+b\right)}
{(\omega+c)^2+a^2}-\frac{\left(\omega+c\right)\sin\left(t(\omega+c)+b\right)
}{\left(\omega+c\right)^2+a^2}\Bigg]\nonumber\\
\fl
&\quad+\frac{1}{2}\frac{\left(\omega-c\right)\sin\left(b\right)}{(\omega-c)^2+a^2}
+\frac{1}{2}\frac{\left(\omega+c\right)\sin\left(b\right)}{(\omega+c)^2+a^2}\nonumber\\
\fl
&\quad+\frac{1}{2}\frac{a\cos\left(b\right)}{(\omega-c)^2+a^2}
-\frac{1}{2}\frac{a\cos\left(b\right)}{(\omega+c)^2+a^2} \,.
\end{eqnarray}

\section*{References}

\bibliography{time-dependent}

\end{document}